\documentclass[12pt, a4paper]{article}
\pdfoutput=1
\usepackage[height=25cm,width=16cm]{geometry}
\usepackage{feynmp}
\usepackage{amsmath}
\usepackage{ascmac}
\usepackage{dcolumn}
\usepackage{bm,here}
\usepackage{subfig}
\usepackage{comment}
\usepackage{ifpdf}
\usepackage{slashed}
\usepackage{colortbl}
\usepackage{color}
\usepackage{comment}
\usepackage{braket}
\usepackage[mathscr]{eucal}
\usepackage[sort&compress, numbers, merge]{natbib}

\usepackage{cancel}

\ifpdf
\usepackage{graphicx}     
\usepackage[bookmarksopen,colorlinks=true,linkcolor=bblue,citecolor=ppink,urlcolor=ppink]{hyperref}
\else     
\usepackage[dvipdfmx]{graphicx}     
\usepackage[dvipdfmx,bookmarksopen,colorlinks=true,linkcolor=bblue,citecolor=ppink,urlcolor=ppink]{hyperref}
\fi

\usepackage{multicol}
\definecolor{red}{rgb}{1,0,0}
\definecolor{ppink}{rgb}{0.921545,0.440586,0.687243}
\definecolor{bblue}{rgb}{0.400000,0.400000,1.000000}
\usepackage[charter]{mathdesign}
\usepackage{soul}
\usepackage{wrapfig}

\newcommand\blfootnote[1]{%
	\begingroup
	\renewcommand\thefootnote{}\footnote{#1}%
	\addtocounter{footnote}{-1}%
	\endgroup
}

\begin{document}
	
\begin{titlepage}
		
\begin{flushright}
    \hfill IPMU21-0013 \\
\end{flushright}
		
\begin{center}

	\vskip 1.5cm
	{\Large \bf Leptophilic fermion WIMP \\ [.3em] $\sim$ Role of future lepton colliders $\sim$}

	\vskip 2.0cm
	{\large Shun-ichi Horigome$^*$\blfootnote{$^*$shunichi.horigome@ipmu.jp},
			Taisuke Katayose$^{\dagger}$\blfootnote{$^\dagger$taisuke.katayose@ipmu.jp}, \\ [.3em]
			Shigeki Matsumoto$^{\ddagger}$\blfootnote{$^\ddagger$shigeki.matsumoto@ipmu.jp}
			and
			Ipsita Saha$^{\$}$\blfootnote{$^{\$}$ipsita.saha@ipmu.jp} }
			
	\vskip 2.0cm
	{\sl Kavli IPMU (WPI), UTIAS, University of Tokyo, Kashiwa, Chiba,  277-8583, Japan} \\ [.3em]
	
    \vskip 3.5cm
    \begin{abstract}
        \noindent
	    The leptophilic weakly interacting massive particle (WIMP) is realized in a minimal renormalizable model scenario where scalar mediators with lepton number establishes the WIMP interaction with the standard model (SM) leptons. We perform a comprehensive analysis for such a WIMP scenario for two distinct cases with an SU(2) doublet or singlet mediator considering all the relevant theoretical, cosmological and experimental constraints at present. We show that the mono-photon search at near-future lepton collider experiments (ILC, FCC-ee, CEPC, etc.) can play a significant role to probe the yet unexplored parameter range allowed by the WIMP relic density constraint. This will complement the search prospect at the near-future hadron collider experiment (HL-LHC). Furthermore, we discuss the combined model scenario including both the doublet and singlet mediator. The combined model is capable of explaining the long-standing muon (g-2) anomaly which is an additional advantage. We demonstrate that the allowed region for anomalous muon (g-2) explanation, which has been updated very recently at Fermi National Accelerator Laboratory, can also be probed at the future colliders which will thus be a simultaneous authentication of the model scenario.
    \end{abstract}
			
\end{center}
		
\end{titlepage}
	
\tableofcontents
\vspace{1cm}
\setcounter{page}{1}
	
\section{Introduction}
\label{sec: intro}

The weakly interacting massive particle (WIMP) has been one of the most accepted particle candidates for dark matter (DM) in the past few decades, for it can naturally explain the DM abundance in the current universe through the standard thermal freeze-out mechanism\,\cite{Bernstein:1985th,Srednicki:1988ce}. The mass of a WIMP particle can vary in between $\mathcal{O}$(1)\,MeV\,\cite{Boehm:2002yz,Boehm:2003bt} and $\mathcal{O}(100)$\,TeV\,\cite{Griest:1989wd,Hamaguchi:2007rb,Hamaguchi:2008rv,Hamaguchi:2009db,Murayama:2009nj,Hambye:2009fg,Antipin:2014qva,Antipin:2015xia,Gross:2018zha,Fukuda:2018ufg}. Specifically, WIMP with mass of the order of the electroweak scale is the most compelling option as plethora of beyond the standard model (BSM) scenario, such as supersymmetric extensions, extra dimensions, etc. cater a valid WIMP candidate along with solving the naturalness problem of the electroweak symmetry breaking. In the general freeze-out scenario, the WIMP candidate is assumed to have interactions with the standard model (SM) particles such that, in the early universe at high temperature, the WIMP was in thermal equilibrium with all the other primordial bath particles. However, the rate of interaction of the DM particle with the SM particle should be enough to explain the current abundance of the DM candidate in the present universe, namely, the {\it relic density}. 

On the other hand, such interaction becomes useful to probe the nature of the WIMP DM at collider and underground (direct detection) experiments. Current DM direct detection and collider experiments, however, relies mostly on the WIMP interaction with the SM quarks, and if the WIMP predominantly interacts with the SM leptons, the current experimental efficiency is insufficient to probe such interaction. Such kind of WIMP with dominant interaction with the SM leptons are generally called as `Leptophilic WIMP'\,\cite{Dev:2013hka,Bai:2014osa,Chang:2014tea,Junius:2019dci,Okawa:2020jea,Kawamura:2020qxo}, and in this article, we will analyze aspects of such a leptophilic WIMP that interacts with the SM leptons via some mediator particles utilizing minimal and renormalizable models.

In general, the WIMP can be classified in terms of their quantum numbers; the spin and the weak isospin quantum numbers\,\cite{Banerjee:2016hsk}. In our study, we will consider a SU(2)$_L$ singlet and spin-half Majorana fermion as our desired leptophilic WIMP particle which is stabilized by an additional $Z_2$ symmetry. Then, it turns out that an additional new particle called the mediator must be introduced to have a renormalizable interaction between the WIMP and the SM leptons, and it must be bosonic in nature. In the minimal renormalizable extension of the SM, one can either have three generations of SU(2)$_L$ doublet scalar mediator corresponding to the left-handed leptons, or three generations of SU(2)$_L$ singlet corresponding to the right-handed SM leptons. The quantum number of the scalar mediators will follow their SM lepton partners, and those must be odd under the new $Z_2$ symmetry. It is to be mentioned that we only focus on lepton flavor conserving interactions of the DM particle.

In view of the above model specification, we first consider two individual model scenarios with either of the doublet and singlet mediators, and perform a comprehensive analysis considering all robust theoretical, cosmological and experimental constraints. The theoretical constraint involves the vacuum stability conditions on the scalar potential with the extra scalar mediators. We carefully discard the parameter space that gives an unstable vacuum by choosing correct scalar quartic couplings. On the other hand, the current experimental constraint includes the relic abundance of the WIMP measured by the Planck collaboration and limits from LEP-II and LHC run II experiments. First of all, the relic abundance of the leptophilic WIMP for the low mass region $\cal O$(GeV) can be achieved with the proper tuning of WIMP-lepton Yukawa couplings. However, at larger WIMP mass of $\cal{O}(\rm TeV)$, the co-annihilation of the WIMP with the scalar mediators will be the key mechanism to obtain the correct relic abundance. In calculating the relic abundance, we have properly taken into account the effect of the SM thermodynamics at the early universe\,\cite{Saikawa:2018rcs, Saikawa:2020swg}. It is to be noted that our scenario consists of a singlet Majorana WIMP and scalar mediators with lepton numbers which resemble the R-parity conserving minimal supersymmetric scenario of the bino-slepton co-annihilation. We thus include the current 95\% C.L. exclusion limit on the bino-slepton mass from LEP-II and latest LHC 13\,TeV run-II results. Additionally, extra constraints coming from the Higgs to di-photon decay mode and from the oblique T-parameter constraint (for the doublet mediator scenario) will also be implemented.

In the next part, we discuss the future prospect of the leptophilic WIMP. Here, the future high-luminosity LHC (HL-LHC) experiment is expected to probe a part of the unexplored parameter region covering a non-degenerate and highly degenerate WIMP-mediator mass region. Thus, we consider future lepton collider experiments as a competent alternative that can probe a mildly degenerate mass region as well. We did a detail analysis for mono-photon searches\footnote{For mono-Higgs searches at the ILC experiment in a similar scenario, please refer to the reference\,\cite{Jueid:2020yfj}.} for the WIMP pair production at the ILC experiment and explore the reach of the searches at 95\% C.L at the ILC-250\,GeV. We find that the ILC can give a complementary detection prospect to the future HL-LHC experiment for such a leptophilic WIMP.

Finally, we discuss the combined model scenario where both the doublet and singlet scalar extensions to the leptophilic WIMP is studied. The model can be preferred over the others as it can also explain the discrepancy between the theory and experiment in the anomalous muon magnetic moment. We find the allowed parameter space that explain both the muon $(g-2)$ and WIMP relic abundance. Finally, we did the same mono-photon search analysis for the combined model, and found that the ILC-250\,GeV can validate the parameter space that can explain the muon $(g-2)$ anomaly. This is an interesting finding as the muon anomaly has persisted also at the Fermilab\,\cite{Abi:2021gix} experiment, the mono-photon search at the ILC will provide an additional clue to the leptophilic WIMP.

The paper is arranged as follows. In section\,\ref{sec: model}, we describe the Lagrangian for the SM gauge singlet Majorana fermionic WIMP and its leptophilic interaction via the scalar mediators explaining all relevant parameters of the potential. We divide sub-sections for each mediator types. Next, in section\,\ref{sec: present}, we discuss all the relevant constraints at the present time starting from the vacuum stability conditions, relic abundance condition and the current constraints from LEP \& LHC experiments. We show the allowed parameter space for each model that satisfy all the current constraints. In section\,\ref{sec: future prospect}, we analyze the current allowed parameter space at the future collider with an emphasis on the mono-photon searches at ILC 250\,GeV. We discuss the complementary between the future hadron and lepton colliders. Finally, in section\,\ref{sec: combined}, we discuss the combined model scenario consisting both the doublet and singlet mediators. Here, we present the calculation for anomalous muon magnetic moment and show the allowed parameter space capable of explaining the experimental anomaly. Additionally, we repeat the mono-photon analysis at the ILC for the combined model and discuss the prospect of simultaneous validation of the model parameter space at the ILC that also explain the anomaly. Lastly, in section\,\ref{sec: conclusion} we summarize our findings and conclude.

\section{Minimal models of the leptophilic WIMP}
\label{sec: model}

We consider minimal and renormalizable models to explore the interaction of a singlet Majorana fermion WIMP that only talks to the SM leptons. In the simplest system composed of the WIMP and the SM particles, no renormalizable interaction exists due to the presence of the SM gauge symmetry and the newly imposed $Z_2$ symmetry making the WIMP stable. The WIMP (SM particles) is charged odd (even) under the $Z_2$. Hence, an additional new particle, called a mediator, is introduced. Possible quantum numbers of the mediator to have a renormalizable interaction between the WIMP and the SM leptons are as follows:\footnote{All the cases lead to the four-point effective interaction $({\bar \chi}\gamma_\mu \gamma_5 \chi)({\bar \ell_i} \gamma^\mu \ell_j)$ with  $\chi$ and $\ell_i$ denoting the Majorana WIMP and the SM leptons with generation index, if the mediator mass is enough greater than the WIMP mass as well as the electroweak scale and integrated out from their original (renormalizable) theories.}
\begin{center}
    \begin{tabular}{r|ccccc}
        Mediator Type & Spin & SU(3)$_c$ & SU(2)$_L$ & U(1)$_Y$ & Z$_2$  \\
        \hline
             Scalar & 0 & {\bf 1} & {\bf 1} & $-1$~~~~ & $-1$ \\
             Scalar & 0 & {\bf 1} & {\bf 2} & $-1/2$   & $-1$ \\
        \hline
             Vector & 1 & {\bf 1} & {\bf 1} & $-1$~~~~ & $-1$ \\
             Vector & 1 & {\bf 1} & {\bf 2} & $-1/2$   & $-1$ \\
             Vector & 1 & {\bf 1} & {\bf 1} & $0$~     & $+1$ \\
        \hline
    \end{tabular}
\end{center}

\noindent
This kind of approach for dark matter models taking full SM gauge symmetry and renormalizability introducing appropriate mediators into account has been first proposed in Ref.\,\cite{Ko:2016zxg}, and it has fixed various problems of dark matter effective field theories as well as dark matter simplified  models that were widely used for a long time to interpret the data on dark matter searches at various experiments and observations. Though their study was mainly focused on interactions between the dark matter and quarks, as also pointed out in the reference, it is straightforward to apply this approach to the leptophilic case.

Renormalizable models including a vector mediator are complicated in general, for it should contain the "Higgs mechanism" to make the mediator massive and a few new (chiral) fermions need to be introduced to make the models anomaly free\,\cite{Madge:2018gfl,Blanco:2019hah}. We therefore focus on the models including scalar mediators in this article; the mediator particle is either a SU(2) scalar doublet with U(1) hypercharge $-1/2$ ($Q = T_3 +Y$) which we named Left-(handed) mediator, or a SU(2) scalar singlet with U(1) hypercharge $-1$, namely Right-(handed) mediator. Analogous to the three generation of the SM leptons, for each case, there are three scalar mediators corresponding to each lepton flavor. For the sake of simplicity, we consider a flavor-universal scenario in this article and define the scalar masses by only one degenerate mass parameter. We explain more on this in the following parts.

On the other hand, in the following sections, we will discuss phenomenology of individual model perspective with Left- or Right-mediators. The effect of introducing both types of the mediators will also be discussed in a later part in Sec.\,\ref{sec: combined} including its motivation.

\subsection{Left-mediator model}
\label{subsec: L-model}

The left-mediator is a SU(2) scalar doublet that has a charged and a neutral components analogous to the left-handed charged-lepton and neutrino in the SM. The complete Lagrangian including the three generations of the left-mediators consists of two additional parts besides the usual SM Lagrangian $({\cal L}_{\rm SM})$ and the kinetic terms of the WIMP and the scalar mediators. These are respectively the interaction between the WIMP and the mediators (denoted by ${\cal L}_{{\rm DM}\,L}$) and the scalar potential (denoted by $V_L$) describing the self-interactions of the new scalar doublets and their interactions with the Higgs boson:
\begin{align}
    & {\cal L}_L = {\cal L}_{\rm SM}
    + \frac{1}{2} {\bar \chi}\left(i\slashed {\partial} - m_\chi \right) \chi
    + (D_L^\mu \tilde{L}_i)^\dagger (D_{L\,\mu} \tilde{L}_i)
    + {\cal L}_{{\rm DM}\,L} - V_L(H, \tilde{L}_i) ,
    \label{eq: L-model} \\
    & {\cal L}_{{\rm DM}\,L} =  - y_L \bar{L}_i \tilde{L}_i \chi  + h.c ,
    \nonumber \\
    & V_L = m_{\tilde{L}}^2\,|\tilde{L}_i|^2  + \frac{\lambda_L}{4}\,|\tilde{L}_i|^4 + \lambda_{LH}\,|\tilde{L}_i|^2 |H|^2
    + \lambda_{LH}^\prime\,(\tilde{L}_i^\dagger \tau^a \tilde{L}_i) (H^\dagger \tau^a H)
    + [\frac{\lambda_{LH}^{\prime \prime}}{4} (\tilde{L}_i^\dagger H^c)^2 + h.c.] .
    \nonumber
\end{align}
where a summation over the repeated indices is implicitly assumed, and the index $i$ spans lepton flavors $(e,\mu,\tau)$. Additionally, $\chi$ is the WIMP, $L_i$ is the SM lepton doublet, $H$ is the Higgs doublet ($H^c \equiv i \sigma_2\,H^*$), $\tilde{L}_i$ is the scalar mediator doublet whose quantum numbers matches with $L_i$, $\tau^a$ is the Pauli matrx and $D_L^\mu$ is the covariant derivative acting on $\tilde{L}_i$.

Since we are assuming lepton flavor universality, the Lagrangian parameters $y_L$, $m_{\tilde{L}}$, $\lambda_L$, $\lambda_{LH}$, $\lambda^\prime_{LH}$ and $\lambda^{\prime \prime}_{LH}$ are common in different flavors. To avoid sizable contribution to tiny neutrino masses, we impose the lepton number symmetry, or to be more precise, the $B-L$ symmetry to the model. This is equivalent to assign the lepton quantum number to the mediator particles, and thus terms proportional to $\lambda_{LH}^{\prime\prime}$ is prohibited. After the electroweak symmetry breaking with $v \simeq 246$\,GeV being the vacuum expectation value of the Higgs field, the physical mass of the each component of the scalar mediator doublets is given as
\begin{align}
    & m_{\tilde{e}_{L_i}}^2 =    (\lambda_{LH} + \lambda_{LH}^\prime)\,\frac{v^2}{2}  + m_{\tilde{L}}^2 ,
    \label{eq: L-masses1} \\
    & m_{\tilde{\nu}_{L_i}}^2 = (\lambda_{LH} - \lambda_{LH}^\prime)\,\frac{v^2}{2}  + m^2_{\tilde{L}} ,
    \label{eq: L-masses2}
\end{align}
with $\tilde{L}_i = (\tilde{\nu}_{L_i}, \tilde{e}_{L_i})^T$. The nomenclature of the physical masses are done in analogy to the superpartners of the leptons in supersymmetric standard models, namely the sleptons.

\subsection{Right-mediator model}
\label{subsec: R-model}

Akin to Left-mediator model, one can write down the full Lagrangian for the right-mediator model where instead of doublets the mediators are SU(2) singlets with the same quantum number as the right-handed SM charged leptons. Explicit form of the Lagrangian is
\begin{align}
    & {\cal L}_R = {\cal L}_{\rm SM}
    + \frac{1}{2} {\bar \chi}\left(i\slashed {\partial} - m_\chi \right) \chi
    + (D_R^\mu \tilde{R}_i)^\dagger (D_{R\,\mu} \tilde{R}_i)
    + {\cal L}_{{\rm DM}\,R} - V_R(H, \tilde{R}_i) ,
    \label{eq: R-model} \\
    & {\cal L}_{{\rm DM}\,R} =  - y_R \bar{E}_i \tilde{R}_i \chi  + h.c ,
    \nonumber \\
    & V_R = m_{\tilde{R}}^2\,|\tilde{R}_i|^2  + \frac{\lambda_R}{4}\,|\tilde{R}_i|^4 + \lambda_{RH}\,|\tilde{R}_i|^2 |H|^2 ,
    \nonumber
\end{align}
where $E_i$ is the SM charged lepton singlet, $\tilde{R}_i$ is the scalar mediator singlet whose quantum numbers resembles $E_i$ and $D_R^\mu$ is the covariant derivative acting on $\tilde{R}_i$. Here also, we consider all the parameters $y_R$, $m_{\tilde{R}}$, $\lambda_R$ and $\lambda_{RH}$ to be the same for different lepton flavors. Taking the same nomenclature as above, physical masses of the right-mediators are given by
\begin{align}
    m_{\tilde{e}_{R_i}}^2 = \lambda_{RH}\,\frac{v^2}{2}  + m^2_{\tilde{R}} .
    \label{eq: R-mass}
\end{align}

\section{Present status of the leptophilic WIMP}	
\label{sec: present}

We discuss all relevant theoretical and experimental constraints imposed to the leptophilic models defined in the previous section, and figure out the present status of the models.

\subsection{Theoretical constraint}
\label{subsec: theoretical constraint}

Since we introduce new scalar fields in the leptophilic WIMP models as the mediators, the stability of the scalar potential has to be confirmed. The complete scalar potential for the left-mediator model includes the SM Higgs potential in addition to $V_L$, and it is given by
\begin{align}
	V^\mathrm{Full}_{L}(H,\,\tilde{L}_i) = \mu^2|H|^2 + \frac{\lambda}{4}|H|^4 + V_L(H,\,\tilde{L}_i) ,
	\label{eq: VFullL}
\end{align}
where the explicit form of the potential $V_L$ is given in eq.\,(\ref{eq: L-model}). First, we obtain the following constraints because of the request that the potential must be bounded from below:
\begin{align}
    \lambda > 0 , \quad \lambda_L > 0 , \quad
    \sqrt{\lambda\,\lambda_L} > 2\,(|\lambda^\prime_{LH}| - \lambda_{LH}) .
    \label{eq: VSL1}
\end{align}
Next, since the $Z_2$ symmetry should not be broken after the electroweak symmetry breaking to make the WIMP stable even at the present universe, the scalar mediators should not develop non-zero vacuum expectation value. Hence, we obtain the other constraint from the request that the masses in eq.\,(\ref{eq: L-masses1}) and eq.\,(\ref{eq: L-masses2}) must be positive with $v = (-4\mu^2/\lambda)^{1/2}$:
\begin{align}
    -\lambda\,m_{\tilde{L}}^2/(2\,\mu^2) > |\lambda^\prime_{LH}|- \lambda_{LH} .
    \label{eq: VSL2}
\end{align}
It is then possible to prove that our vacuum, namely the potential minimum with the vacuum expectation value of the Higgs field being $\langle H \rangle = (0, v/\sqrt{2})^T$ while that of the scalar mediator being $\langle \tilde{L} \rangle = 0$, becomes a global one when the constraints (\ref{eq: VSL1}) and (\ref{eq: VSL2}) are satisfied.

Similar to the above, the constraints for  a stable vacuum in the right-mediator model is
\begin{align}
    \lambda > 0 , \quad \lambda_R > 0 , \quad
    \sqrt{\lambda\,\lambda_R} > -2\lambda_{RH} , \quad
    -\lambda\,m_{\tilde{R}}^2/(2\,\mu^2) > -\lambda_{RH} ,
    \label{eq:VSR}
\end{align}
which also make the vacuum stable as the global minimum of the scalar potential.

\subsection{Astrophysical and cosmological constraints}
\label{subsec: AC consgtraints}
	
\subsubsection{Relic abundance}
\label{subsubsec: relic abundance}

In the early universe, the WIMP was in thermal equilibrium with all the SM particles residing in the thermal bath, and its abundance is determined by the so-called 'freeze-out' mechanism. The amount of thermal relic of the WIMP can be theoretically estimated by solving the Boltzmann equation. On the other hand, the mean mass-density of the cold dark matter content of the present universe is measured by the Planck experiment that observes the cosmic microwave background as well as by other astrophysical observations\,\cite{Aghanim:2018eyx}:
\begin{align}
    \Omega h^2 = 0.120 \pm 0.001 .
\end{align}
Weak-scale interaction for thermal WIMPs can easily obtain the correct relic abundance with a thermally averaged annihilation cross-section into SM particles of $\langle \sigma v \rangle = \mathcal{O}(1)$\,pb.

\begin{figure}[t]
    \centering
    \includegraphics[width=30mm,height=30mm]{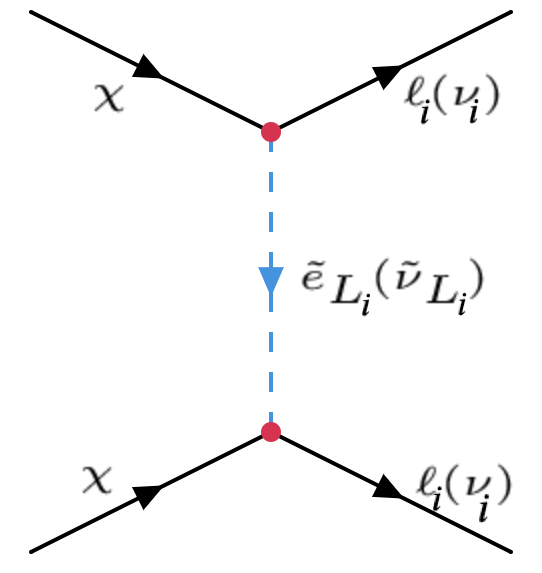}
    \includegraphics[width=30mm,height=30mm]{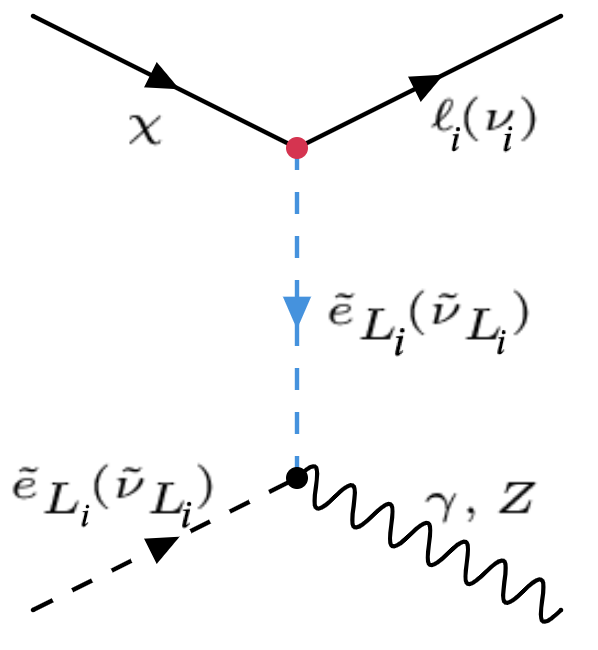}
    \includegraphics[width=30mm,height=30mm]{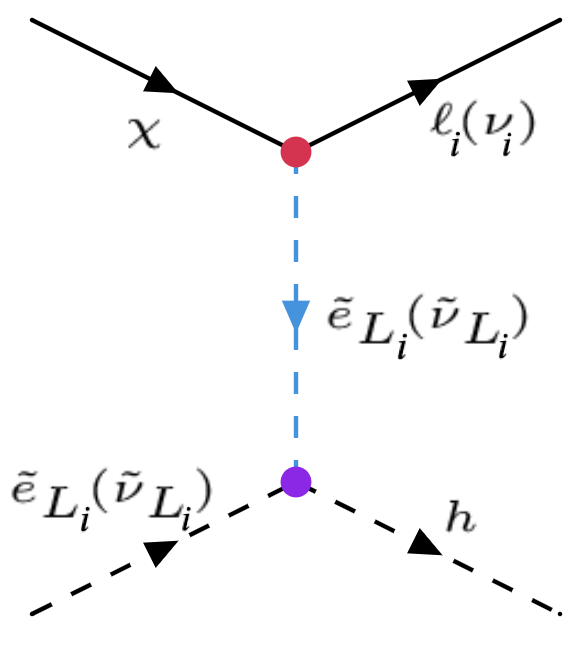}
    \includegraphics[width=35mm,height=30mm]{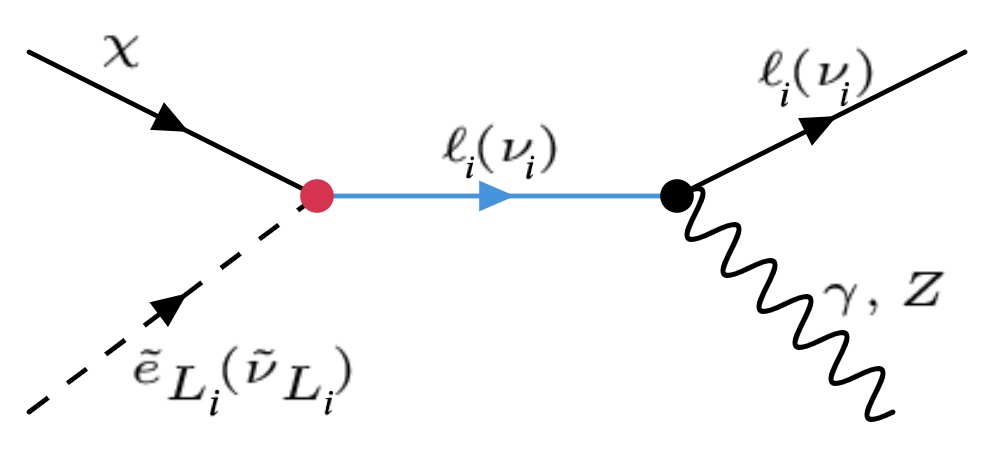}
    \\
    \qquad\qquad\qquad
    \includegraphics[width=30mm,height=30mm]{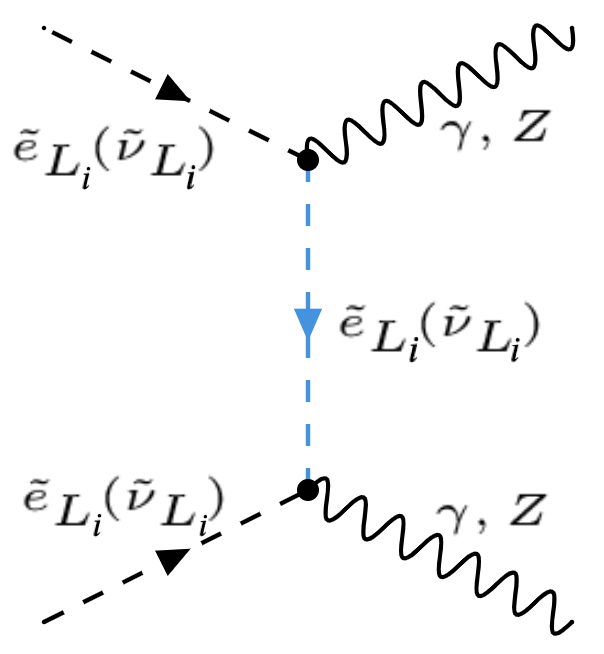}
    \includegraphics[width=30mm,height=30mm]{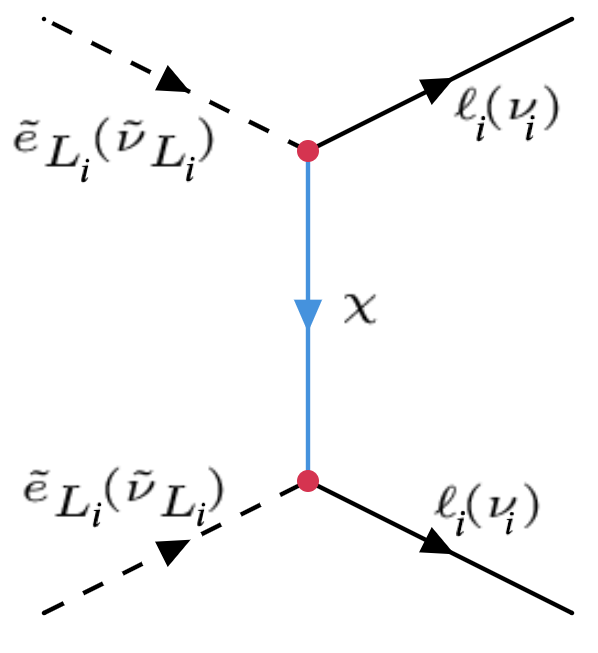} 
    \includegraphics[width=30mm,height=30mm]{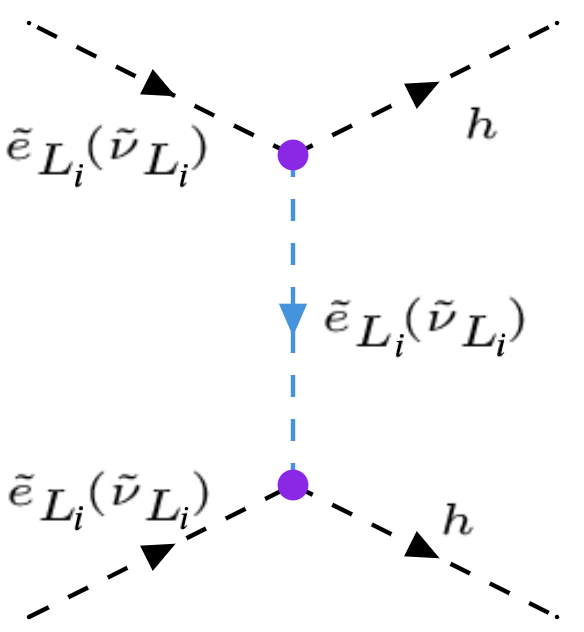} 
    \caption{\small \sl Some illustrative Feynman diagrams for self-annihilation and co-annihilation processes to calculate the relic abundance of the WIMP in the left-mediator model. Red colored vertices corresponds to Yukawa coupling $y_L$, while purple vertices are proportional to the scalar coupling $\lambda_{LH}$. For the right mediator model, $\tilde{e}_{L_i}$ should be replaced by $\tilde{e}_{R_i}$ and there will be no neutral partner.}
    \label{fig: FD}
\end{figure}

In our case, when the scalar mediators are much heavier than the WIMP, the leptophilic WIMP only annihilates into the SM leptons and contributes to the relic abundance, as shown by the top-left diagram in Fig.\,\ref{fig: FD} for the left-mediator model. In this limit, the WIMP annihilation cross section only depends on the WIMP mass $(m_\chi)$, the mediator mass ($m_{\tilde{e}_{L_i}}$ \& $m_{\tilde{\nu}_{L_i}}$ or $m_{\tilde{e}_{R_i}}$) and the Yukawa coupling ($y_L$ or $y_R$) for the left- or right-mediator model, respectively. On the other hand, when the mediator mass and the WIMP mass are degenerate within 10\%, the relic density is controlled by co-annihilation processes\,\cite{Griest:1990kh}. The scalar quartic couplings between the mediators and the SM Higgs also become important in this limit for the relic density calculation, as shown by several diagrams in Fig.\,\ref{fig: FD} for the left-mediator model. In our analysis, we have scanned over the model parameter space considering both the limits that yield the correct relic abundance by the WIMP self-annihilation and the co-annihilation processes. The relic density is calculated utilizing the code {\bf micrOMEGAs-v5}\,\cite{Belanger:2018mqt}. 

Uncertainty at the relic density calculation due to SM thermodynamics has recently been estimated\,\cite{Saikawa:2018rcs, Saikawa:2020swg}, where it is shown that the uncertainty of the effective massless degrees of freedom can be as large as 10\%, depending on the temperature of the universe. This in turn induces $\cal O$(5\%) uncertainty to the WIMP density during the freeze-out. We have incorporated this uncertainty in the {\bf micrOMEGAs} using the data provided in ref.\,\cite{Saikawa:2020swg}.

\subsubsection{Direct and Indirect detections}

Direct dark matter detection is known to be very powerful to search for various WIMP candidates. The detection, however, relies on the (coherent) scattering between WIMP and nucleus, so that it is not efficient for the leptophilic WIMPs. Here, one might think that such a scattering emerges radiatively through one-loop diagrams even in the leptophilic WIMP models we are discussing, where the SM leptons are propagating in the loop. The scattering cross section of such a process, however, turns out to be too small to detect at the present and near future detectors when the leptophilic WIMP is a Majorana fermion\,\cite{Baker:2018uox}. Hence, we do not include any constraints from the direct WIMP detection in our analysis. 

On the other hand, the indirect dark matter detection is known to be very useful to search for WIMP candidates irrespective of their interactions, where the signal strength is proportional to the WIMP self-annihilation cross section. The annihilation cross section in the leptophilic models we are discussing is, however, velocity (p-wave) suppressed due to the angular momentum and CP conservation, and is insignificant at the present universe. Therefore, the indirect WIMP detection constraints are also irrelevant to our scenario.

\subsection{Collider constraints}

\subsubsection{Direct mediator productions at LHC and LEP experiments}
\label{subsubsec: LHC and LEP}

It is difficult to directly probe the WIMP at the current LHC experiment, since it is a gauge singlet and interacts only with the SM leptons. On the other hand, the mediator particles are charged under SM gauge interactions and accessible both at lepton and hadron colliders. To evaluate constraints from the collider experiments on the leptophilic models, we can directly use those for supersymmetric particles. This is because the leptophilic models become the same as the ($R$-parity conserving) minimal supersymmetric model (MSSM) with the lightest and next-lightest supersymmetric particles being the bino and the (degenerate) sleptons, respectively, while other sparticles decoupled, when we fix the parameters of the leptophilic models to be $y_L = g'/\sqrt{2}$, $\lambda_L = (g^2 + g^{\prime 2})/2$, $\lambda_{LH} = g^{\prime 2} \cos (2\beta)$, $\lambda'_{LH} = g^2 \cos (2\beta)/4$, $\lambda''_{LH} = 0$, and $y_R = \sqrt{2} g'$, $\lambda_R = 2 g^{\prime 2}$, $\lambda_{RH} = -2 g^{\prime 2} \cos (2\beta)$.\footnote{Since the MSSM has two Higgs doublets to make the model free from quantum anomaly, we take the decoupling limit at its Higgs sector to make a comparison between the MSSM and the leptophilic models.} Here, $g$, $g'$ are the SU(2)$_L$ and U(1)$_Y$ gauge couplings, respectively, while $\tan \beta$ is the ratio of the vacuum expectation values of the two Higgs doublets in the MSSM. On the other hand, the sleptons are produced by electroweak Drell-Yan processes at the collider experiments and decays into the bino and a SM lepton with 100\% branching fraction. Hence, the signal strength does not depend on Yukawa and scalar interactions addressed above, and we can directly use collider limits on the supersymmetric particles to evaluate collider constraints on the leptophilic models.

\begin{figure}[t]
    \centering
    \includegraphics[width=49mm]{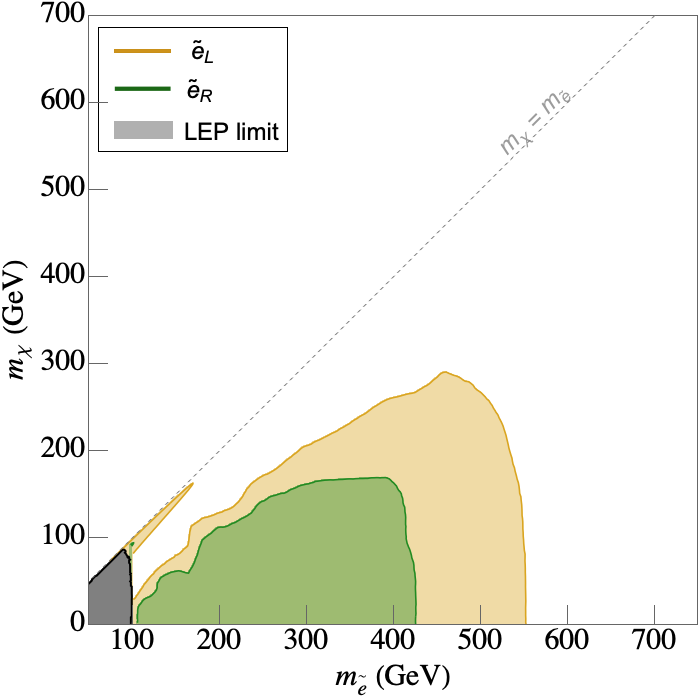}
    \quad
    \includegraphics[width=49mm]{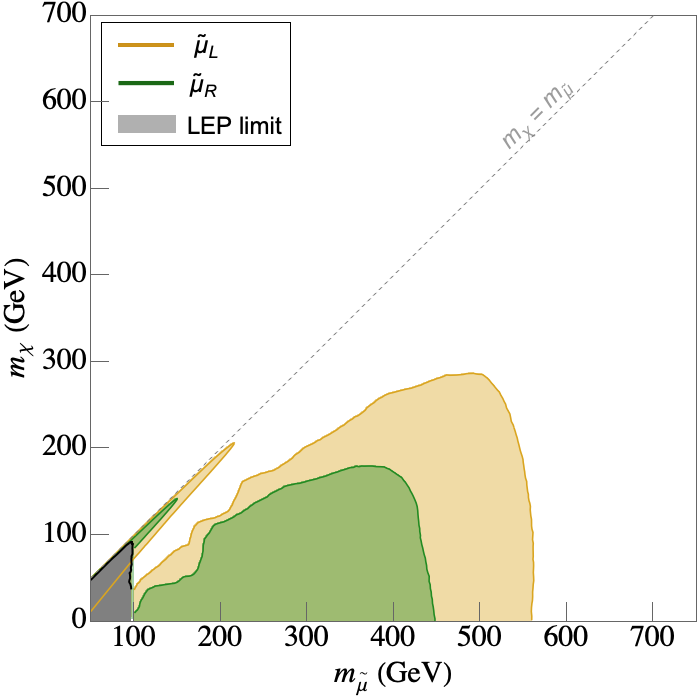}
        \quad
    \includegraphics[width=49mm]{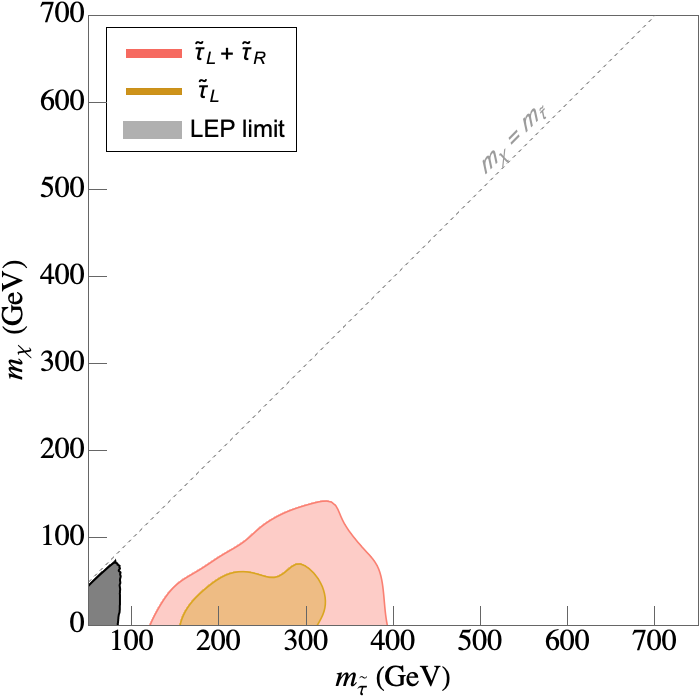}
    \caption{\small \sl The 95\% C.L. exclusion (observed) limit on the mediator-WIMP mass plane obtained from the analysis of various slepton pair productions at the LHC (8 \& 13\,TeV) and LEP-II  experiments.}
    \label{fig: collider}
\end{figure}

The LEP experiment has searched for supersymmetric (electrically) charged sleptons which decays dominantly into a charged SM lepton and a bino-like lightest supersymmetric particle (a Majorana fermion)\,\cite{LEPII}. As a result, the LEP experiment excluded the right-handed smuon, the super partner of the right-handed muon, with its mass below 94\,GeV for a neutralino-smuon mass gap above 10\,GeV\,\cite{pdg2020}. This is a model-independent bound on leptonic charged scalars in our case, as mentioned above. We therefore impose this bound on the right-mediators and, as a conservative limit, also on the left-mediators. The excluded region by LEP for each charged slepton is also shown as the gray-shaded area in Fig.\,\ref{fig: collider}.

The LEP experiment has also searched for supersymmetric neutral sleptons, i.e. sneutrinos, where each sneutrino decays into a neutrino and the lightest neutralino. The model independent bound on the mass of such a sneutrino is from the invisible decay width search for the $Z$ boson, as discussed at the last part of this subsection. On the other hand, it is also possible to search for the sneutrinos using the mono-photon channel caused by its  pair production associated with a photon\,\cite{Zyla:2020zbs}. However, in our models, the Drell-Yan process via the $s$-channel exchange of the $Z$ boson is the only process contributing to the mono-photon signal, i.e. the other process via the $t$-channel exchange of the chargino does not contribute to the signal. As a result, no additional constraint on the sneutrino mass, which is severer than the model independent bound mentioned above, is obtained at the LEP experiment due to the smallness of the signal cross section\,\cite{Datta:2002jh}. Hence, we do not include the constraint that is obtained by the sneutrino pair production associated with a photon in our analysis.

The LHC experiment has also looked for such simplified scenarios where charged sleptons with 100\% branching ratio into its SM lepton partner and a WIMP (a bino-like LSP) can be produced. After the Run-II of the LHC experiment at 13\,TeV center of mass energy, the ATLAS collaboration has reported the 95\% exclusion limit on both left and right-handed sleptons (selectrons and smuons) at 139\,fb$^{-1}$ luminosity for slepton-neutralino mass difference more than 80\,GeV\,\cite{Aad:2019vnb}. Dedicated search for compressed spectra where the slepton-neutralino mass difference become as low as 550\,MeV for a slepton mass around 70\,GeV has also been done and the 95\% C.L. (confidence level) exclusion limit is presented\,\cite{Aad:2019qnd}. In left and middle panel of Fig.\,\ref{fig: collider}, we show this excluded region for the selectron and smuon pair productions as yellow and green shaded regions for the left and right-handed sleptons, respectively. The earlier limit from 8\,TeV run of the LHC experiment has also been included in the contours. For comparison purposes, we also show the excluded region for stau productions (left-handed stau production and left + right-handed stau production)\,\cite{Aad:2019byo, Sirunyan:2019mlu} in the right panel of Fig.\,\ref{fig: collider}. As seen from the excluded region for the stau, the constraints from the stau searches are weaker than those of selectron and smuon. Here, we would like to remind us that we only consider the degenerate, flavor universal case, and therefore we use the most-sensitive smuon search limit on our model scenarios as the strongest bound.

Neutral sleptons, i.e. sneutrinos, have also been produced at the LHC experiment by the electroweak Drell-Yan process, and the experiment is possible to give some constraints on the sneutrinos if their decays produce enough amount of visible particles\,\cite{Zyla:2020zbs}. The sneutrinos in our models, however, decay invisibly, i.e. each sneutrino decays into a neutrino and the dark matter, at 100\,\% branching ratio, and the possible way to search for the sneutrinos is the so-called mono-X search with $X$ being a jet, photon, etc. Such signals have been intensively studied in Ref.\,\cite{Balazs:2017ple}, and no severer constraint than that from the invisible $Z$ boson decay width search is obtained. Hence, we do not include such a constraint in our analysis.

\subsubsection{Radiative correction to the Higgs decay into diphoton}
\label{subsubsec: diphoton}

\begin{figure}[t]
    \centering
    \includegraphics[width=73mm]{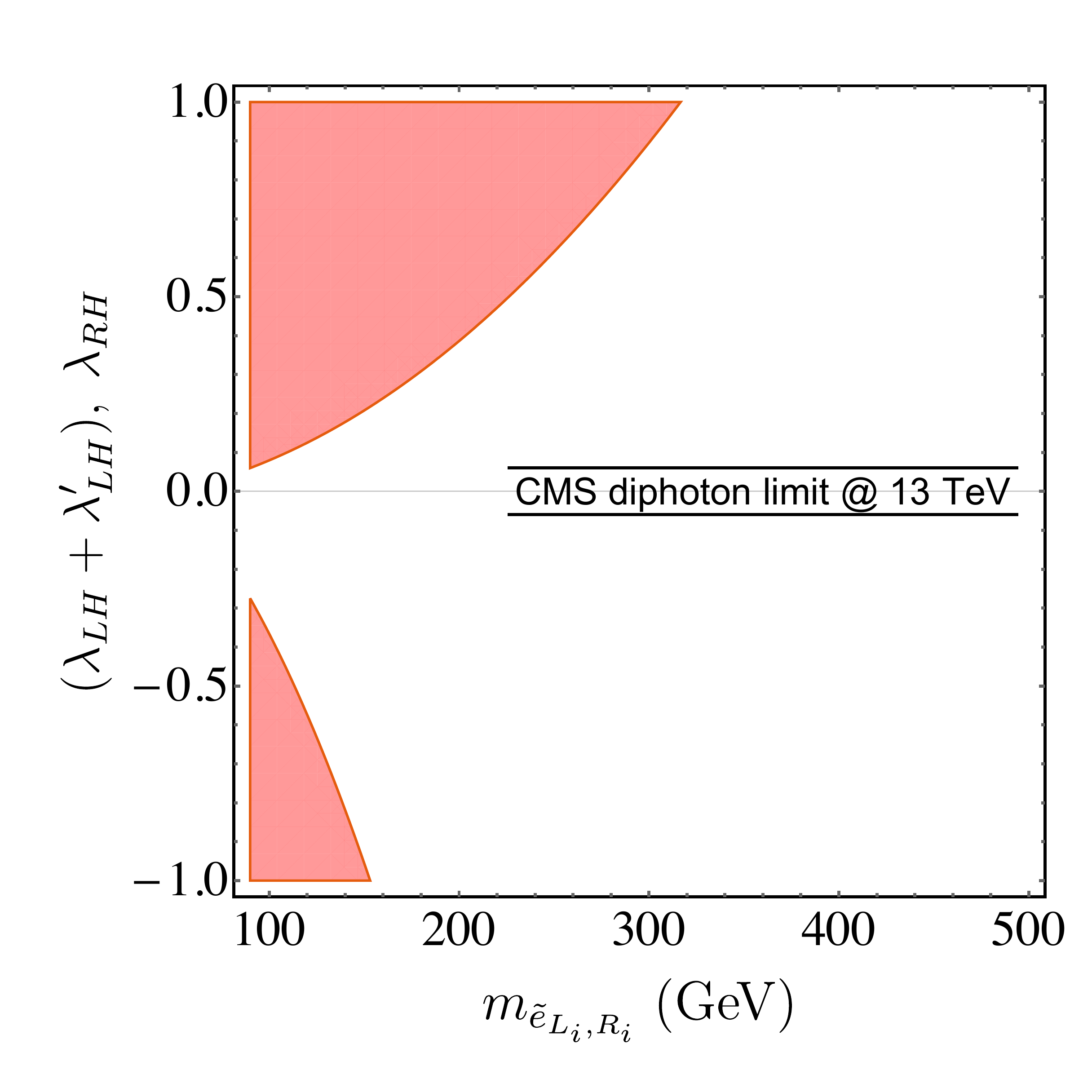}
    \qquad
    \includegraphics[width=73mm]{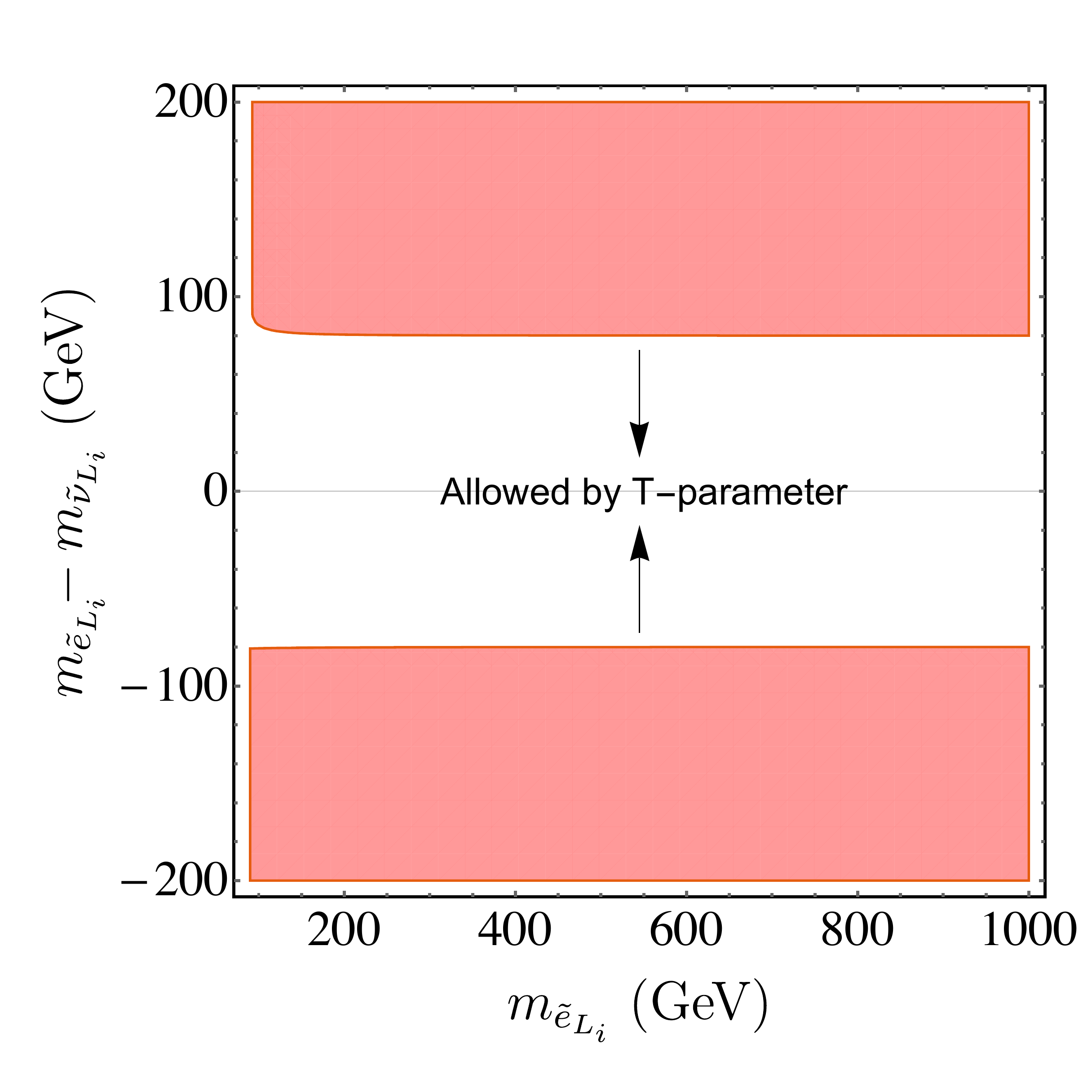}
    \caption{\small \sl The white region depicts the allowed parameter space at 95\% C.L. limit from current Higgs to di-photon decay width measurement (left) and oblique T-parameter constraint (right).}
    \label{fig: Tparam Diphoton}
\end{figure}

With the accumulation of more data, the Higgs data at the LHC experiment has been updated with unprecedented accuracy and it shows an increasing affinity to the SM value. In our model, the tree-level Higgs decay branching will exactly follow the SM value. However, the left and right charged mediator can significantly contribute to the loop induced Higgs to diphoton decay\,\cite{Gunion:1989we,Djouadi:2005gi,Djouadi:2005gj}. The latest constraint on the Higgs to diphoton signal strength is given by the CMS collaboration as $\mu_{\gamma\gamma} = 1.18^{+0.17}_{-0.14}$~\cite{Sirunyan:2018ouh}. The signal strength is defined as the ratio of the Higgs production cross-section times its branching ratio to the gamma gamma mode with the corresponding SM value. Since the exotic charged mediator does not contribute to the production channel and considering that the total decay width only changes negligibly due to the diphoton mode, the signal strength in our case can be approximated as the ratio between the partial decay width of the Higgs to diphoton decay to its SM value. Now, charged mediator coupling to the SM Higgs is given by the scalar quartic couplings $(\lambda_{LH} + \lambda_{LH}^{\prime})$ for the left-mediator and $\lambda_{RH}$ for the right-mediator\,\cite{Djouadi:1996pb}. In Fig.\,\ref{fig: Tparam Diphoton}, we show the excluded parameter space in the charged mediator mass vs its coupling to the Higgs at the 95\% C.L. as the red-shaded region. We have included this constraint in our analysis.

\subsubsection{Electroweak precision measurement}
\label{subsec: Tparam}

In the left-mediator case, the scalar mediators are doublet under the SU(2)$_L$ gauge group. Therefore, those can significantly contribute to the self-energy of the SM electroweak gauge bosons, and an additional constraint from the so-called oblique T-parameter value that combines the electroweak precision data should be considered\,\cite{He:2001tp,Grimus:2007if,Grimus:2008nb}. The contribution to the T-parameter from each generation of the left-mediator particles is given as follows\,\cite{Barbieri:2006dq}:
\begin{align}
    \Delta T_{\tilde{L}_i} &= \frac{1}{16 \pi^2 \alpha v^2}
    \left(
        \frac{ m_{\tilde{e}_{L_i}}^2 + m_{\tilde{\nu}_{L_i}}^2 } {2}
        - \frac{ m_{\tilde{e}_{L_i}}^2 m^2_{\tilde{\nu}_{L_i}} } { m_{\tilde{e}_{L_i}}^2 - m_{\tilde{\nu}_{L_i}}^2 }
        \log{ \frac{ m_{\tilde{e}_{L_i}}^2 }{ m_{\tilde{\nu}_{L_i}}^2 } }
    \right) \nonumber \\
    &=
    \frac{( m_{\tilde{e}_{L_i}} - m_{\tilde{\nu}_{L_i}} )^2}{16\pi^2 \alpha v^2}
    \left[
        \frac{2}{3}
        - \frac{ ( m_{\tilde{e}_{L_i}} - m_{\tilde{\nu}_{L_i}} )^2 }{ 30 m_{\tilde{\nu}_{L_i}}^2 }
        + {\cal O} \left( \left\{ \frac{m_{\tilde{e}_{L_i}} - m_{\tilde{\nu}_{L_i}}} {m_{\tilde{\nu}_{L_i}}}
         \right\}^4 \right)
    \right] ,
    \label{eq: Tparam}
\end{align}
where $\alpha$ is the fine structure constant. We see that, when the charged and neutral mediator particles are nearly degenerate in mass, the T-parameter depends not on their absolute masses but on the mass difference between them at leading order. Since, we have three generations of degenerate scalar mediators, all of them will contribute to the T-parameter and put a stringent constraint on the mass splitting between the charged and the neutral mediator particles. In the right panel of Fig.\ref{fig: Tparam Diphoton}, we show the 95\% exclusion limit on the charged vs the charged to neutral mass difference plane, where the latest value of the T-parameter constraint from new physics, $\Delta T = 0.05 \pm 0.06$ is used.\,\cite{pdg2020}. As seen in the figure, one can not get a mass gap larger than 80\,GeV for the charged scalar mass of interest. 

\subsubsection{Invisible Higgs and Z boson decays}

\begin{figure}[t]
    \centering
    \includegraphics[width=75mm]{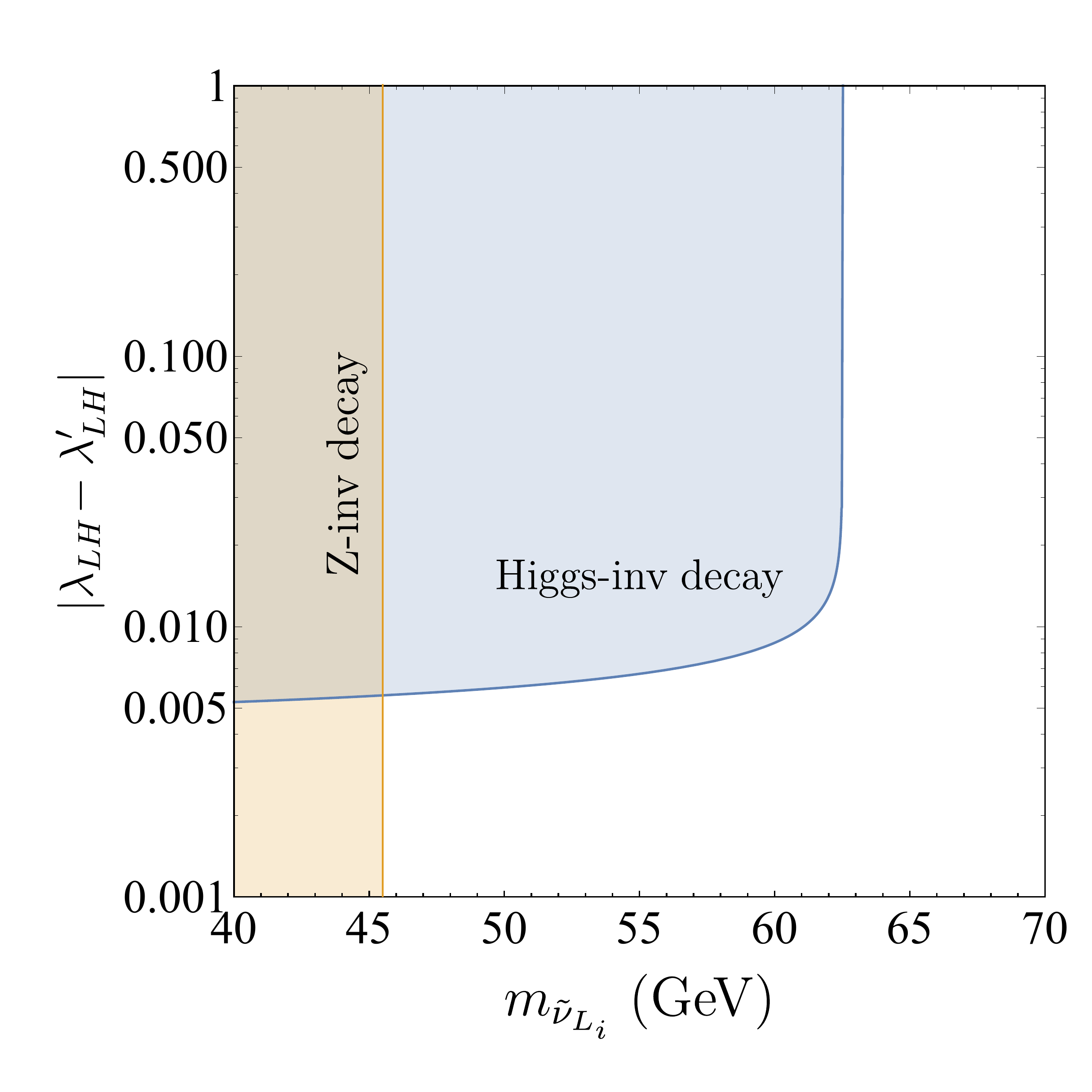}
    \caption{\small \sl Constraints from the invisible decay width search of the Higgs boson at the present LHC experiment (blue shaded) and that of the $Z$ boson at the LEP experiment (orange shaded).}
    \label{fig: invisible}
\end{figure}

We have another constraint on the left mediators from precision Higgs and $Z$ boson invisible decay measurements. Although the LEP constraint demands the charged components of the left mediators to be, at least, greater than 90\,GeV, there is no severe constraint on the mass of the neutral components. If the mass of neutral components is smaller than half of Higgs or $Z$ boson mass, there is a decay channel such as $h \to \tilde{\nu}_{L_i} \tilde{\nu}_{L_i}^* \to \chi \chi \nu_i \bar{\nu}_i$ or $Z \to \tilde{\nu}_{L_i} \tilde{\nu}_{L_i}^* \to \chi \chi \nu_i \bar{\nu}_i$. These processes contribute to the invisible decay width of the Higgs or Z boson as
\begin{eqnarray}
   \Gamma(h \to \tilde{\nu}_{L_i} \tilde{\nu}_{L_i}^*) &=& \frac{(\lambda_{LH}-\lambda^\prime_{LH})^2 v^2}{16 \pi m_h} \sqrt{1-\frac{4m^2_{\tilde{\nu}_{L_i}}}{m^2_h}} , \\
   \Gamma(Z \to \tilde{\nu}_{L_i} \tilde{\nu}_{L_i}^*) &=& \frac{\alpha\, m_Z}{48 \sin^2 \theta_W \cos^2 \theta_W}\left(1-\frac{4m^2_{\tilde{\nu}_{L_i}}}{m^2_Z}\right)^{3/2} ,
\end{eqnarray}
where $m_h$ and $m_Z$ are the Higgs and $Z$ boson masses, respectively, $\alpha$ is the fine structure constant, and $\theta_W$ is the Weinberg angle. The invisible decay width of the $Z$ boson is usually expressed using the effective neutrino number coupling to the $Z$ boson and the observation at the LEP experiment presently gives its number to be $N_\nu = 2.9840 \pm 0.0082$\,\cite{ALEPH:2005ab} at present, which restricts the mass of the neutral component to be larger than half of the $Z$ boson mass $(m_{\tilde{\nu}_L} \geq m_Z/2)$. On the other hand, the invisible decay width of the Higgs boson, or in other words, the invisible decay branching fraction of the Higgs boson which is defined by
\begin{eqnarray}
     {\cal B}_{\rm inv} = \frac{3 \Gamma(h \to \tilde{\nu}_{L_i} \tilde{\nu}_{L_i}^*)}{\Gamma(h \to {\rm SMs}) + 3\Gamma(h \to \tilde{\nu}_{L_i} \tilde{\nu}_{L_i}^*) } ,
 \end{eqnarray}
with $\Gamma(h \to {\rm SMs}) \simeq 4.1$\,MeV \cite{Asner:2013psa} is constrained to be ${\cal B}_{\rm inv} \leq 0.13$ by ATLAS collaboration at the LHC experiment\,\cite{ATLAS:2020cjb}. The constraints from invisible decays of Higgs and $Z$ bosons are summarized in Fig.\,\ref{fig: invisible} on the ($m_{\tilde{\nu}_{L_i}}, \lambda_{LH}$)-plane, and we include those in our analysis.

\subsection{Constraints from anomalous magnetic moments of leptons}
\label{subsec: g-2}

As we will discuss in more details in Sec.\ref{sec: g-2}, the leptophilic WIMP models inherently predict the contribution to anomalous magnetic moments of leptons such as muon and electron. At present, experimental data concerning the moments shows us the following result: 
\begin{align}
    \Delta a_\mu \equiv a^\mathrm{exp}_\mu - a^\mathrm{SM}_\mu &= 251\,(59) \times 10^{-11}\, \text{\cite{Abi:2021gix}}
    \label{eq: muon g-2 discrepancy} \\
    \Delta a_e \equiv a^\mathrm{exp}_e - a^\mathrm{SM}_e &= -88\,(36) \times 10^{-14}\, \text{\cite{Hanneke:2008tm,Parker_2018}},
    \label{eq: electron g-2 discrepancy}
\end{align}
in terms of the deviation from the SM prediction. As we will also show quantitatively in Sec.\ref{sec: g-2}, both the left- and right-mediator models predict negligible contributions to the anomalous magnetic moments, i.e. those are smaller than the size of the errors shown above. It means that the both models are difficult to explain the anomalous magnetic moment of the muon (if we take its 4$\sigma$ discrepancy between the experimental result and the SM prediction seriously). On the other hand, the discrepancy for the anomalous magnetic moment of the electron is around 2$\sigma$ level and thus it should be used to constrain the models. The contributions to the moment of the electron from the models are, however, negligibly small as addressed, so that we do not have serious constraints from the $\Delta a_e$ result.

\subsection{Present status of the leptophilic WIMP}
\label{subsec: Present status}

Following all the constraints on the previous subsections, we set the range of scan for the model parameters. First of all, we restrict that the mediator mass is greater than the WIMP mass, i.e. $ m_{\chi} < \{m_{\tilde{e}_{L_i},},m_{\tilde{\nu}_{L_i}},m_{\tilde{e}_{R_i}}\}$, and fix the quartic parameters $\lambda_{L}$, $\lambda_{R} = 1$ which do not contribute to any physical observables and only required to be positive from the vacuum stability conditions given in section\,\ref{subsec: theoretical constraint}. The other relevant parameters are varied as
\begin{align}
    &
    |\lambda_{LH}| \leq 1 , \quad
    | y_L| \leq 1 , \quad
    90\,\mathrm{GeV} \leq m_{\tilde{e}_{L_i}} \leq 2\,\mathrm{TeV} , \quad
        |m_{\tilde{e}_{L_i}} - m_{\tilde{\nu}_{L_i}}| \leq 80\,\mathrm{GeV} , \quad  m_{\tilde{\nu}_{L_i}} > 45\,\mathrm{GeV} ,
    \nonumber \\
    &
    | \lambda_{RH} | \leq 1 , \quad
    | y_R | \leq 1 , \quad
    90\,\mathrm{GeV} \leq m_{\tilde{e}_{R_i}} \leq 2\,\mathrm{TeV} , \quad
    1\,\mathrm{GeV} \leq m_\chi \leq 2\,\mathrm{TeV} .
    \label{eq: scan ranges}
\end{align}
Note that, the mass gap between the two components of the SU(2) doublet in the left mediator model takes care of the oblique T-parameter constraint in Fig.\,\ref{fig: Tparam Diphoton}. For collider, Higgs to diphoton and to invisible constraints, we directly use 2$\sigma$ exclusion contours on our scanned parameter space in the allowed region of the correct relic abundance of the WIMP.

Next, we discuss our findings in detail. We scan our parameter space in the range given in eq.\,(\ref{eq: scan ranges}) incorporating constraints on the scalar quartic couplings from Higgs to diphoton and to invisible width searches shown in Fig.\,\ref{fig: Tparam Diphoton} as well as the theoretical constraint in section\,\ref{subsec: theoretical constraint}. The scanning is done with the help of the tool {\fontfamily{qcr}\selectfont emcee}\,\cite{2013PASP..125..306F} to make a proper sampling. The allowed parameters are then passed through the micrOMEGAs code to perform the relic abundance calculation. It is worth mentioning here that the experimental uncertainty of the dark matter relic abundance is much weaker than the uncertainty in the theoretical calculation that comes from the massless degrees of freedom in the early universe. Therefore, to find out the allowed parameter space by the relic abundance constraint, we did a $\chi^2$-analysis considering the 2$\sigma$ uncertainty in the  theoretical calculation and choosing the central value at $\Omega h^2 = 0.120$ from the PLANCK experimental data.

\begin{figure}[t]
    \centering
    \includegraphics[width=73mm]{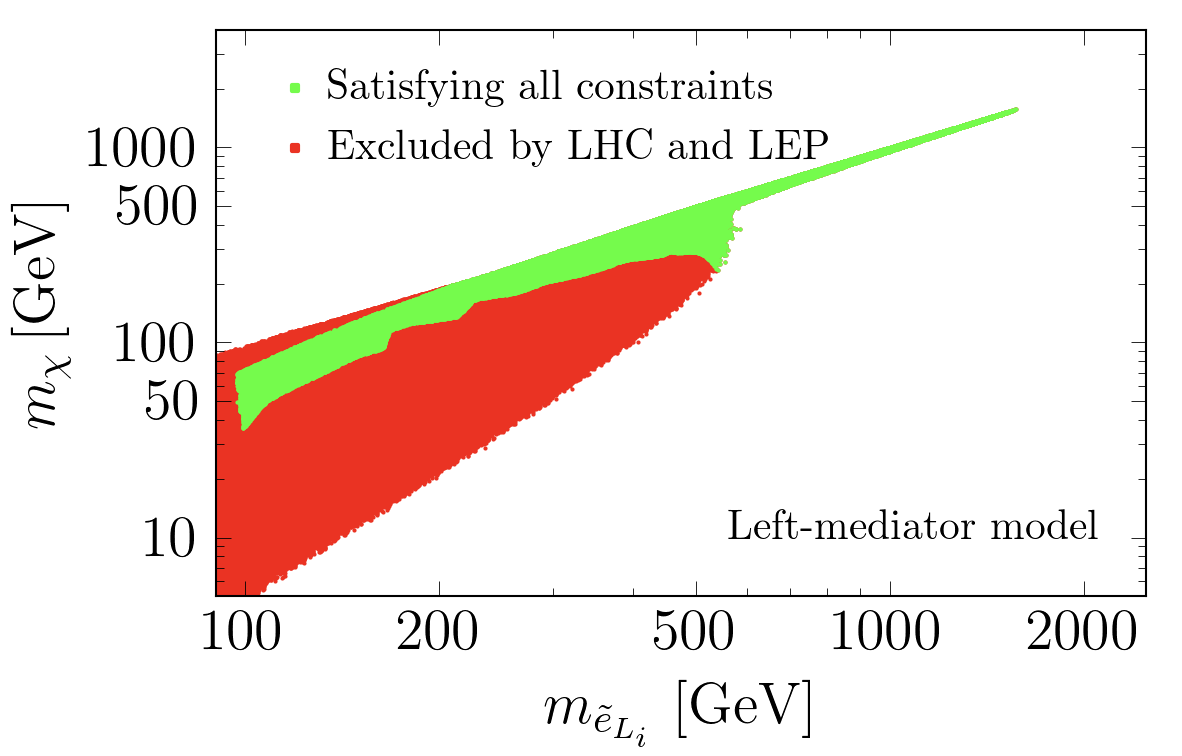}
    \qquad
    \includegraphics[width=73mm]{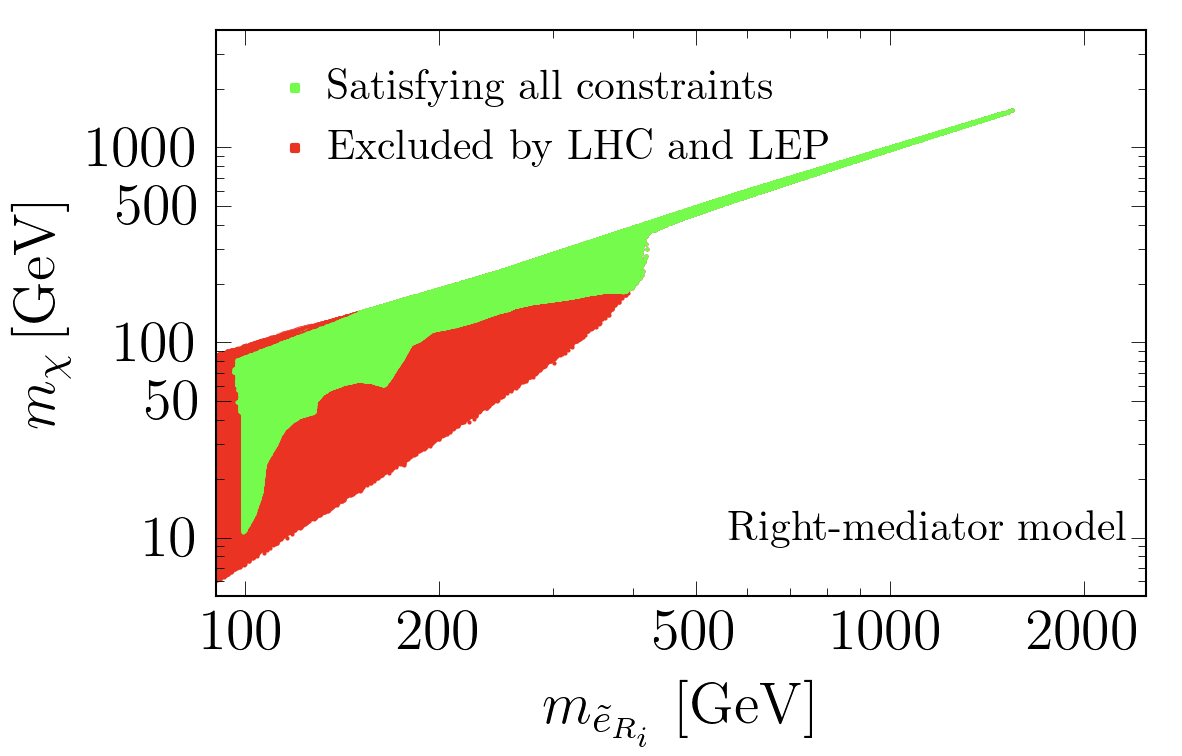} \\
    \includegraphics[width=73mm]{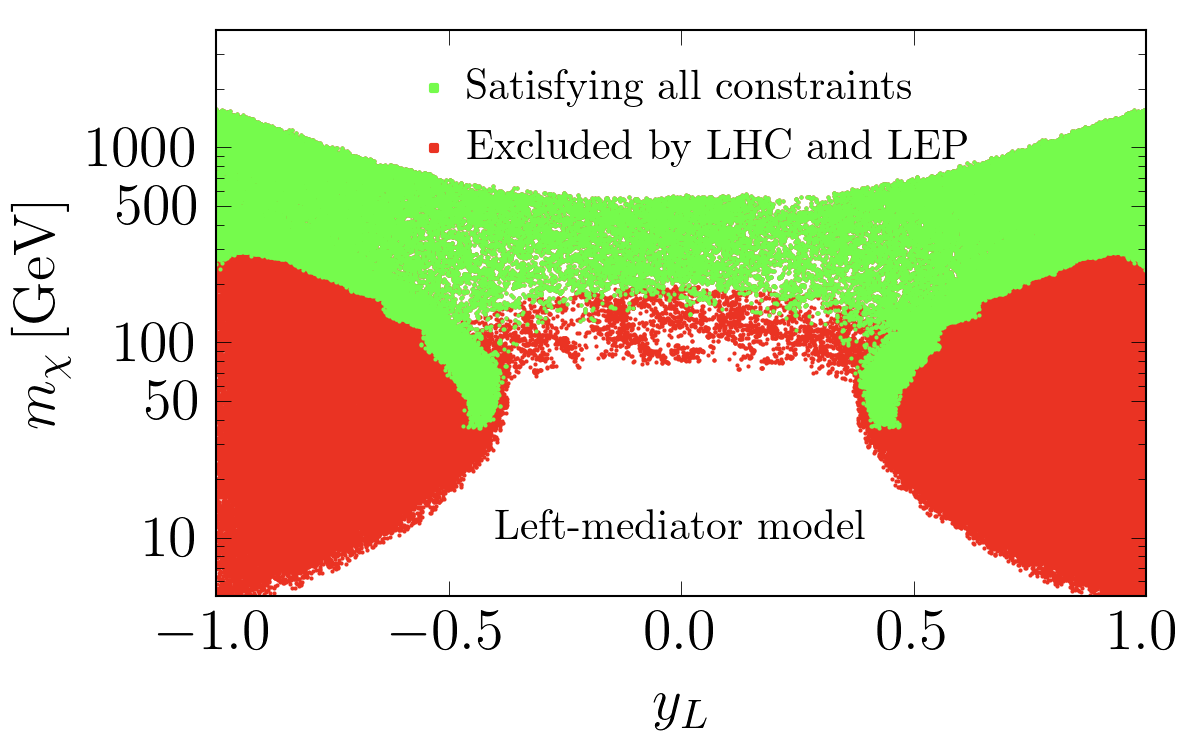}
    \qquad
    \includegraphics[width=73mm]{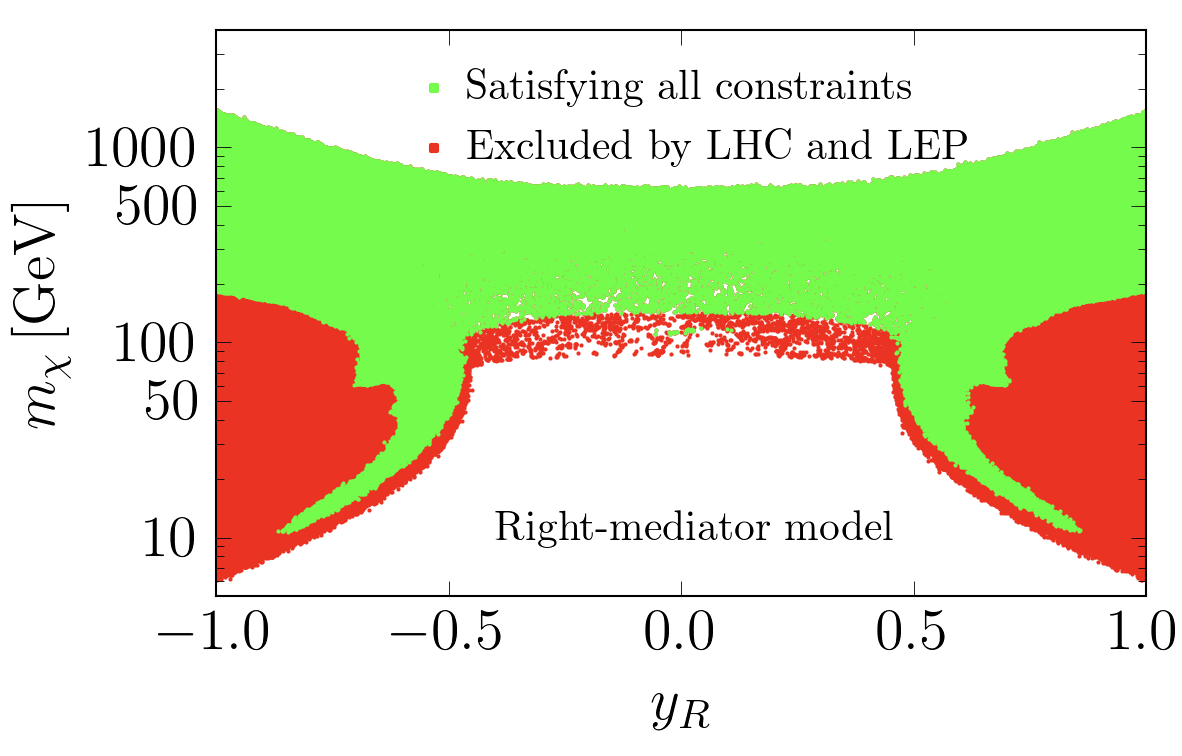}
    \caption{\small \sl Allowed parameter space at 95\% C.L. from theoretical and experimental constraints in the $(m_\chi, m_{\tilde{e}_{L_i}\,(\tilde{e}_{R_i})})$-plane (top panels) and in the $(y_L\,(y_R), m_{\chi})$-plane (bottom panels) for the left-mediator (left panels) and right-mediator (right panels) models, respectively. See text for details.}
    \label{fig: present status}
\end{figure}

In Fig.\,\ref{fig: present status}, we show the presently allowed parameter space on the $(m_{\tilde{e}_{L_i}\,(\tilde{e}_{R_i})}, m_{\chi})$-plane for the left-mediator (top-left panel) and right-mediator (top-right panel) models. The region spread by green points are allowed by the present constraints at 95\% C.L. It is evident from the figure that, at large WIMP mass, only degenerate masses for the WIMP and the mediators can satisfy the correct relic abundance via the co-annihilation mechanism. In this co-annihilating mass region, the scalar quartic coupling of the mediators becomes effective and the correct relic abundance for the largest WIMP mass is achieved for the largest value of the scalar coupling $\lambda_{LH\,(RH)} \sim 1$ which renders an upper bound on the co-annihilating mass range at 1.5\,TeV. On the other hand, at small WIMP mass, the Yukawa coupling needs to be large enough to keep the annihilation cross section around 1\,pb. This is because the WIMP and the mediators are still required to be (mildly) degenerate in mass, so that the annihilation cross section is proportional to $\sim y_L^4\,(y_R^4)/m_\chi^2$ (with $m_\chi \sim m_{\tilde{e}_{L_i}(\tilde{e}_{R_i})})$ unless the co-annihilation comes into play. This fact can also be seen in the bottom panels of Fig.\,\ref{fig: present status}, where the presently allowed parameter space on the $(y_L\,(y_R), m_{\chi})$-plane is shown. In the left-mediator model, the co-annihilation takes the dominant role when the WIMP mass is $m_\chi > 500\,{\rm GeV}$, while, for the right-mediator case, it appears beyond 400\,GeV. The presence of an additional degree of freedom ($\tilde{ \nu}_{L_i}$) in the left-mediator model helps in allowing larger WIMP mass with the correct relic abundance via the WIMP self-annihilation process.

Furthermore, the region spread by red points in all the panels of the figure depicts the region excluded only by the direct searches of the scalar mediators at the LHC and LEP experiments discussed in section\,\ref{subsubsec: LHC and LEP}. As expected, a part of the self-annihilation region has already been excluded by the searches, while degenerate mass regions survive and will be important to be probed at future lepton colliders, as discussed in the following section.

\section{Future prospects of the leptophilic WIMP}  
\label{sec: future prospect}

In this section, we discuss the sensitivity to search for the leptophilic WIMP at future collider experiments, in particular, focusing on the International Linear Collider (ILC), within the L- and R-mediator models, and figure out future prospects of the leptophilic WIMP search.

\subsection{Expected sensitivity at the future colliders}

\subsubsection{Radiative correction to the Higgs decay into diphoton}

\begin{figure}[t]
    \centering
    \includegraphics[width=75mm]{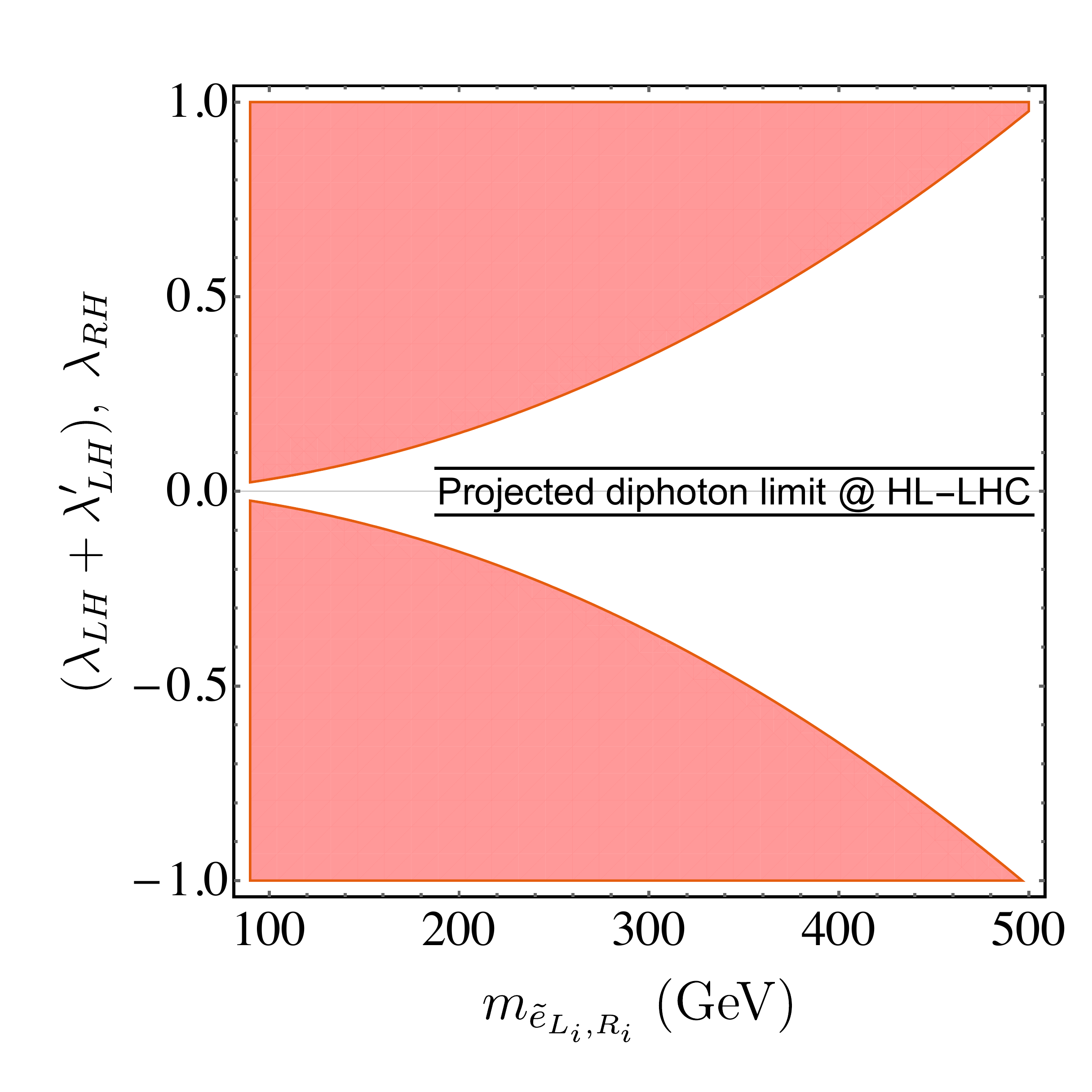}
    \qquad
    \includegraphics[width=75mm]{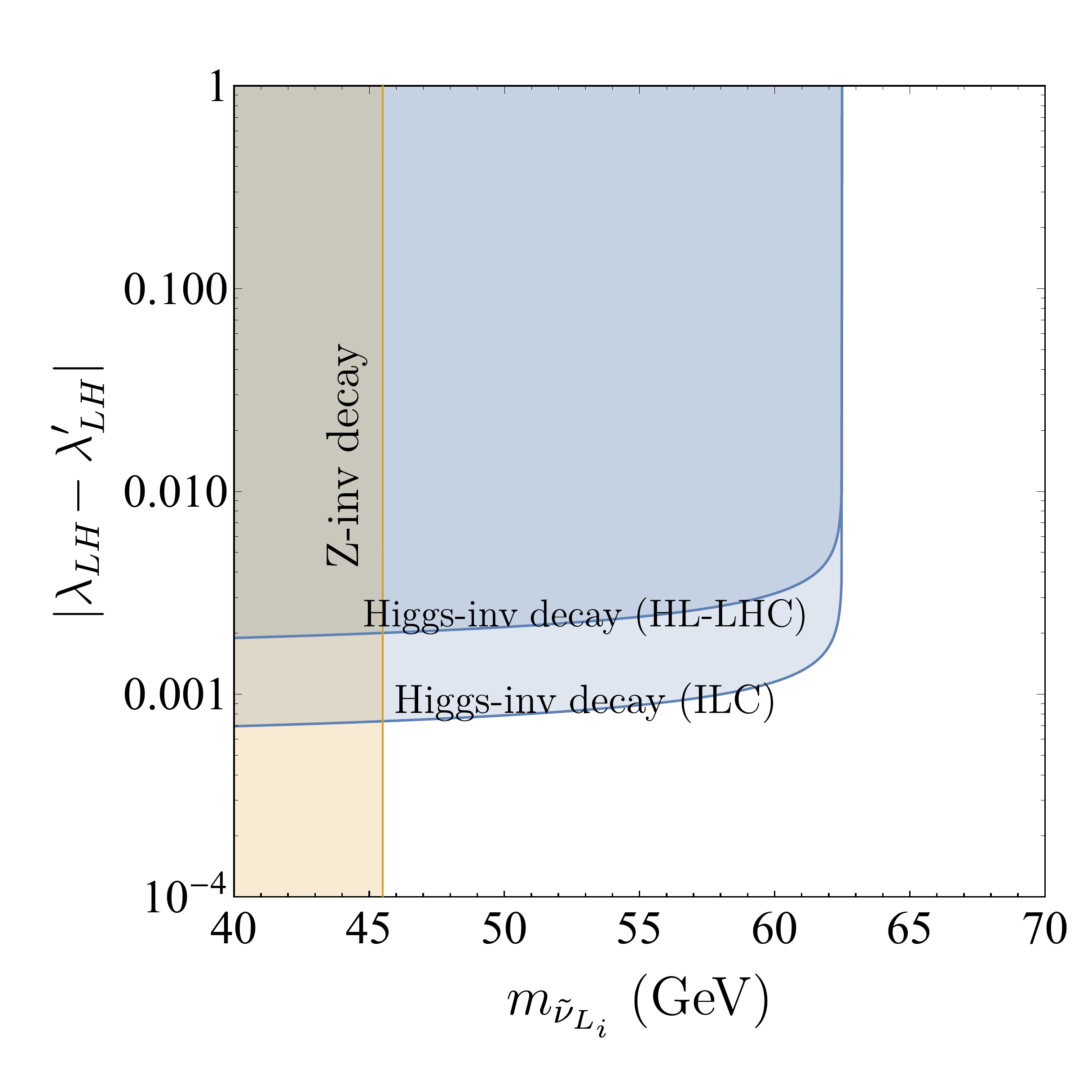}
    \caption{\small \sl The shaded region depicts the projected reach of parameter space at 95\% C.L. from the Higgs to di-photon (left) and to invisible (right) decay width measurements. See text for details.}
    \label{fig: futurediphotn}
\end{figure}

Future projection of the Higgs to diphoton decay mode predicts that it will reach an accuracy at around 2\% level at the HL-LHC\,\cite{Cepeda:2019klc}. Therefore, the coupling of the charged scalar mediators to the Higgs will be further searched for. In Fig.\,\ref{fig: futurediphotn}, we show the projected sensitivity from the Higgs to diphoton searches at the HL-LHC experiment assuming that the uncertainty will be reduced to 2\% while the central value being equal to the SM value.

\subsubsection{Invisible Higgs boson decay}

The measurement of the invisible Higgs decay width will be updated at the HL-LHC experiment in the near future, and it is expected to be further improved at the ILC experiment. Assuming that no new physics signal is observed at the measurements, the constraint on the branching fraction of the invisible Higgs decay width will be obtained as\,\cite{deBlas:2019rxi}
\begin{eqnarray}
    {\cal B}_{\rm inv} \leq
    \begin{cases}
        0.019  & {\rm [HL\mathchar`-LHC]}, \\
        0.0026 & {\rm [250\,GeV \; ILC]}.
    \end{cases}
\end{eqnarray}
With these expected limits, the projected sensitivities of measuring the invisible Higgs decay width at the HL-LHC and ILC experiments are shown in Fig.\,\ref{fig: futurediphotn} on the $(m_{\tilde{\nu}_L}, \lambda_{LH})$-plane.

\subsubsection{Direct mediator productions at the HL-LHC experiment}
\label{subsec: HL-LHC}

There are presently no projected reaches that have been reported by the HL-LHC working group for direct slepton pair productions except for those of stau productions\,\cite{ATLAS:2018diz}, which are shown in Fig.\ref{fig: future prospects} as (blue \& orange) thin lines (95\% C.L. expected reach). These two lines (named case\,1 \& 2) correspond to results of two different analyses; one assumes that systematic uncertainties at the HL-LHC experiment (14\,TeV \& 3\,ab$^{-1}$) will decrease compared to the LHC Run-2 experiment (case 1), while the other assumes that the uncertainties at the HL-LHC experiment are the same as those at the Run-2 experiment (case 2). Projected reaches for selectron and smuon pair productions are currently not available and those are expected to be better than the stau pair productions, though it is not trivial how efficiently the searches cover the degenerate mass region between the mediator and dark matter.

\subsubsection{Direct mediator productions at the ILC experiment}
\label{subsubsec: ilc}

Direct pair productions of charged mediators at the ILC experiment are intensively studied in Ref.\,\cite{Baum:2020gjj} within the context of SUSY models. Among various charged mediators, the production of mediators associated with the muon, i.e. smuon-type mediators, offers the most sensitive search for both the left- and right-mediator models. According to the reference, it turns out that the mass of the mediators less than 250\,GeV, i.e. $m_{\tilde{e}_{L_2/R_2}} \leq 125$\,GeV, is covered at the ILC experiment with 250\,GeV center of mass energy and 500\,fb$^{-1}$ luminosity.

\subsubsection{Mono-photon search at the ILC experiment}
\label{subsubsec: ilc mono-photon}

When the WIMP mass is lower than half of the center of mass energy at the ILC experiment, the mono-photon ($\gamma$) channel works effectively because relatively large Yukawa couplings $y_{L/R}$ are required to achieve the observed relic density. However, when the WIMP and the mediators are degenerate, it become difficult to search for the WIMP using the mono-$\gamma$ channel because the couplings can be small due to the co-annihilation mechanism. In such a case, we can use the direct pair productions of charged mediators addressed above and the channel with pair creation of neutral mediator particles with an additional high energy photon. In the latter case, the mediator will finally decay into a neutrino and a WIMP, and thus it can be counted as the mono-$\gamma$'s. The cross section of this channel is not suppressed by the Yukawa couplings because of the existence of the Drell-Yan process, and we will have a severe constraint also on the co-annihilation region. The main background against the mono-photon signals comes from the mono-$\gamma$ with the pair creation of neutrinos, and we calculated cross sections of the signal and the background for each energy bin by integrating analytic formulae of the differential cross sections. In order to suppress other backgrounds, we set the conditions $E_\gamma \geq$ 10\,GeV and $\cos{\theta_\gamma} \leq$ 0.98, where $E_\gamma$ and ${\theta_\gamma}$ are the energy and the polar angle of the photon, following the similar prescription as in reference\,\cite{Ghosh:2019rtj}.

\subsection{Future prospects of the leptophilic WIMP}

\begin{figure}[t]
    \centering
    \includegraphics[width=73mm]{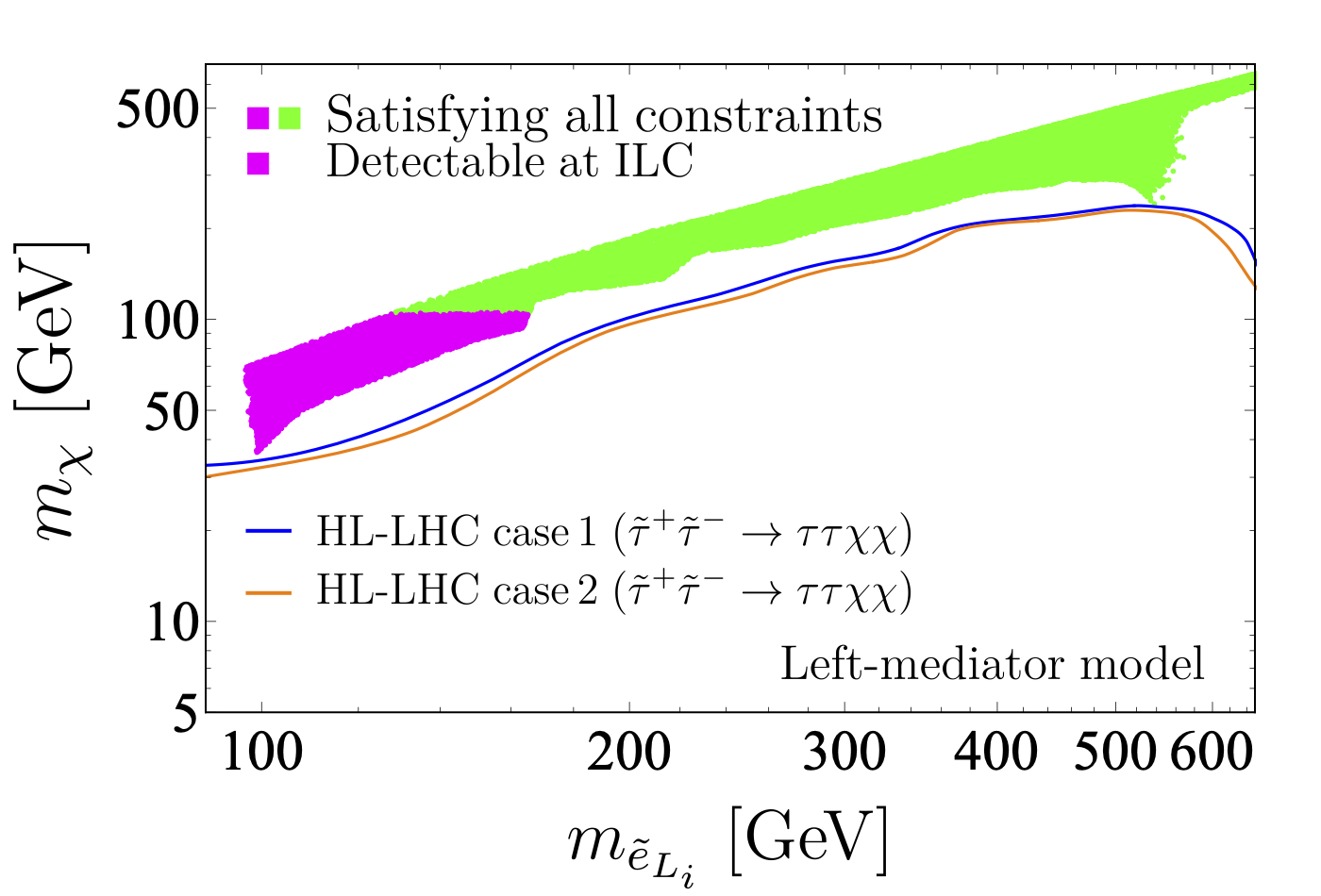}
    \qquad
    \includegraphics[width=73mm]{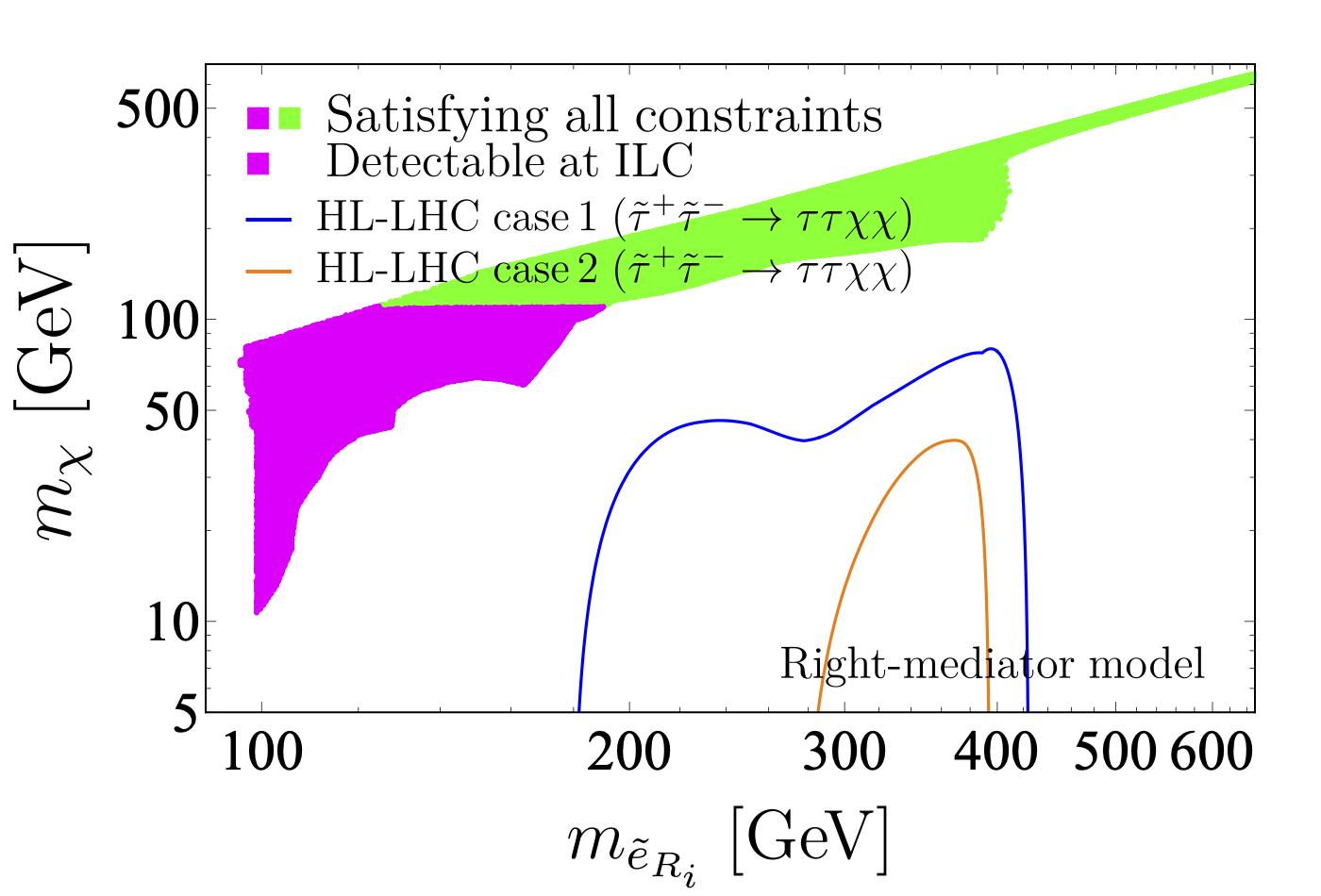} 
    \caption{\small \sl Allowed parameter space for the left and right mediator models is shown in the left and right panels, respectively. The region spread by points is allowed by all the constraints discussed in the previous section, while that spread by magenta points is the projected 95\% C.L. reach at the 250\,GeV ILC experiment taking both the mediator and mono-photon searches into account. Blue and orange lines are expected 95\% C.L. reach of the direct stau pair production at the HL-LHC experiment based on two assumptions concerning systematic uncertainties. See the text for more detail.}
    \label{fig: future prospects}
\end{figure}

We show the projected reach of the ILC experiment on the mediator-WIMP mass plane for left and right mediator models in the left and right panels of Fig.\,\ref{fig: future prospects}, respectively. The region spread by green and magenta points is the allowed parameter space set by all the constraints discussed in the previous section, namely it is the same as the region spread by the green points in Fig.\,\ref{fig: present status}. On the other hand, the improvement of the Higgs to diphoton and invisible decay width measurements in future at the HL-LHC and the ILC experiments does not change the allowed parameter space of the leptophilic WIMP on the $(m_{\tilde{e}_{L_i}}, m_{\chi})$-plane, and thus the benefit of the measurements for the WIMP search is not seen in the figure. The region spread by magenta points in the figure is the 95\% C.L expected reach of the ILC experiment obtained by the search of direct charged mediator productions and that of the mono-photon. To perform the signal-background analysis for the mono-photon search, we followed the approach of our previous study\,\cite{Ghosh:2019rtj}:
\footnote{Our approach reproduced the result of the analysis for the mono-photon search at the LEP experiment\,\cite{Fox:2011fx} with good accuracy\,\cite{Ghosh:2019rtj}, justifying the 
approach to evaluate the expected sensitivity of the ILC experiment.} First, we have analytically calculated the differential cross sections of the signal and the background processes ($e^- e^+ \to \chi \chi \gamma$, $\tilde{\nu}_{L_i} \tilde{\nu}_{L_i}^\ast \gamma$, and $\nu \bar{\nu} \gamma$). Next, taking the initial state radiation\,\cite{Kuraev:1985hb}, beamstrahlung\,\cite{Datta:2005gm} and detector effects\,\cite{Behnke:2013xla} into account, we have numerically integrated the cross sections to obtain signal and background events in each energy bin with respect to the kinematical selection mentioned in section\,\ref{subsubsec: ilc}.
\footnote{The bin width is set to be 5\,GeV within in the energy range of the emitted photon, $E_\gamma$, with $E_\gamma \geq 10$\,GeV.} Finally, we have performed a simple likelihood analysis by comparing these signal and background predictions with the expected experimental data assuming 250\,GeV center of mass energy, 500\,fb$^{-1}$ luminosity and 0.1\,\% systematic uncertainty. We have not utilized the polarization option of the ILC experiment, i.e. we have considered an unpolarized electron and positron collision, for the signal is strong enough and the sensitivity of the experiment to search for the leptophilic WIMP is determined by the center of mass energy of the collision with almost irrespective of the polarization.

The figure shows that the projected reach of the leptophilic WIMP mass at 95\% C.L obtained by the mono-photon search at the ILC experiment is around 110\,GeV. As we have shown in Sec.\,\ref{subsec: HL-LHC}, the HL-LHC experiment is expected to have a good sensitivity to prove a larger mediator mass region, which is less degenerate with the WIMP. On the other hand, in the presently allowed parameter space with the electroweak scale mediator, which is nothing but the region that the ILC experiment will probe, we see more or less a degeneracy in mass between the mediator and the WIMP. The ILC and HL-LHC experiments are thus expected to play complementary roles with each other to search for the leptophilic WIMP.

\section{Combined left- and right-handed model}
\label{sec: combined}

In this section, we discuss the consequences of considering the combined mediator scenario, namely the leptophilic WIMP model having both the right- and left-handed mediators. After addressing the minimal model (Lagrangian) and constraints on it briefly, we discuss the motivation of the scenario and figure out the role of the future colliders to test the model.

\subsection{The minimal model}
\label{subsec: combined model}

\subsubsection{Lagrangian}
\label{subsubsec: combined model}

The renormalizable Lagrangian which contains both the mediators is given as follows:
\begin{align}
    & {\cal L}_{LR} = {\cal L}_{\rm SM}
    + \frac{1}{2} {\bar \chi}\left(i\slashed {\partial} - m_\chi \right) \chi
    + (D_L^\mu \tilde{L}_i)^\dagger (D_{L\,\mu} \tilde{L}_i)
    + (D_R^\mu \tilde{R}_i)^\dagger (D_{R\,\mu} \tilde{R}_i) 
    + {\cal L}_{\rm DM} 
    - V_{LR}(H, \tilde{L}_i,\tilde{R}_i) ,
    \nonumber \\
    & {\cal L}_{\rm DM} = {\cal L}_{{\rm DM}\,L} + {\cal L}_{{\rm DM}\,R} ,
    \nonumber \\
    & V_{LR} =  V_L(H, \tilde{L}_i) + V_R(H, \tilde{R}_i)
    + \lambda_{LR} |\tilde{L}_i|^2 |\tilde{R}_i|^2 + (A\,m_i \tilde{L}_i^\dagger H \tilde{R}_i + h.c.) .
    \label{eq: combined model}
\end{align}
`A' is a dimensionless parameter contributing to the tri-linear coupling and $m_{i}$ is the mass of the SM leptons ($e$, $\mu$, $\tau$). This is an non-trivial assumption on the tri-linear scalar couplings that is inspired by some relations from SUSY models or that similar to the minimal flavor violation\,\cite{DAmbrosio:2002vsn, Colangelo:2008qp}. As seen in the following discussion, it enables us to explain the experimental anomaly of the muon anomalous magnetic moment while avoiding the constraint from that of the electron in the combined model. Moreover, though the assumption makes the tri-linear couplings suppressed, we can find the parameter region that satisfies the relic abundance condition and explains the muon $(g - 2)$ anomaly, as we will also see later. In other words, if we relax the assumption, we will find a wider parameter region that explain both the dark matter relic abundance and the muon anomaly with heavier dark matter and mediator masses. Other parameters in the above lagrangian are the same as those in section\,\ref{sec: model}.

\subsubsection{Vacuum stability constraint}
\label{subsubsec: VSC at combined model}

After the electroweak symmetry breaking, the `A' term causes the mixing between $\tilde{L}_i$ and  $\tilde{R}_i$. We scale this tri-linear coupling by the lepton mass parameters as discussed in section\,\ref{subsubsec: combined model}, so that it is most sensitive to the third generation mixing. From the condition that the diagonalized mass of the mediators must be positive in our vacuum, we have
\begin{align}
    |\,A\,v\,m_\tau| \leq \sqrt{2}\,m_{\tilde{e}_{L_3}} m_{\tilde{e}_{R_3}} .
    \label{eq: local minimum condition for CM}
\end{align}
Remember the fact that the mediator masses $m_{\tilde{e}_{L_i}}$ and $m_{\tilde{e}_{R_i}}$ must be greater than the electroweak scale due to the collider constraints in section\,\ref{subsubsec: LHC and LEP}, the above constraint is always satisfied whenever the `A' term is ${\cal O}(1-10)$. On the other hand, in order to guarantee the stability of our vacuum, we should confirm that the vacuum is the global minimum of the potential or its lifetime is enough longer than the age of the universe. We have hence utilized the result in Ref.\,\cite{Duan:2018cgb} to put a constraint on the `A' term, where the condition of the vacuum stability is carefully investigated up to one-loop level in the framework of MSSM.\footnote{In order to make this vacuum stability constraint more accurate and also confirm the asymptotic safety of the combined model in eq.\,(\ref{eq: combined model}), it is better to go beyond the one-loop calculation by taking into account the renormalization group running of the couplings (model parameters)\,\cite{Hiller:2019mou, Hiller:2020fbu}. On the other hand, the one-loop calculation is justified assuming the combined model is defined at the energy scale that is not much away from the electroweak scale and the coupling constants are below ${\cal O}(1)$ as we adopted in this paper.} According to the result, the constraint is given by $|A| \leq A_c$ with $A_c \simeq 80 + 0.3\,(m_{\tilde{e}_{L_3}}/[{\rm GeV}])$, when $m_{\tilde{e}_{L_3}} = m_{\tilde{e}_{R_3}}$ and the dark matter is degenerate with the mediators in mass to some extent such as a surviving parameter region shown in Fig.\ref{fig: combined model}. We thus adopt this constraint to guarantee the stability of our vacuum in the analysis of the combined model.\footnote{Honestly speaking, this constraint can be relaxed in the combined model defined in eq.\,(\ref{eq: combined model}), as the quartic couplings of the mediators are taken to be large (without conflicting with any constrains discussed so far) in the combined model, while those are fixed by the gauge couplings in the MSSM. Note, however, that our discussion does not change even if we adopt the tighter constraint mentioned in the main text, as we are focusing on the parameter region that the future lepton collider experiments can access at their first stages.}

\subsubsection{Relic abundance constraint}
\label{subsubsec: RAC at combined model}

As we have discussed above, we are interested in the parameter region where all the WIMP and the mediators are in the electroweak scale, so that, for the sake of simplicity, we scan the model parameters assuming $m_{\tilde{e}_L} = m_{\tilde{e}_R}$ and $y_R^4 = 2 y_L^4$. The latter assumption is obtained by requiring $\sigma v(\chi\chi \to e_R \bar{e}_R) = \sigma v(\chi\chi \to e_L \bar{e}_L) + \sigma v(\chi\chi \to \nu_L \bar{\nu}_L)$, namely both the mediators equally contribute to the relic abundance of the WIMP when both the mediators have the same mass, unless coannihilation processes come into play. This simplified analysis is enough in order to show the capability of the future lepton colliders to search for the attractive parameter region of the combined model motivated by the muon $(g-2)$ anomaly as well as the dark matter abundance (thermal relic scenario). Uncertainties at the relic density calculation are taken into account in the same manner as those in section\,\ref{subsubsec: relic abundance}.

\subsubsection{Other constraints}

Whenever the `A' term is ${\cal O}(1-10)$, all other constraints discussed in the previous sections can be applied to the combined model with a good approximation. For constraints from the direct mediator and WIMP production at present and future collider experiments, where the first and second generation mediators play an essential role to provide sensitive signals, the effect of the `A' term is negligibly small. For those from branching fraction measurements of the Higgs decay into diphoton, Higgs and $Z$ boson invisible decays, and also the electroweak precision measurement, the existence of the `A' term affects their signal strengths sub-dominantly. The direct dark matter detection at underground experiments is as ever irrelevant to the leptophilic WIMP even within the combined model. The signal strength of the indirect dark matter detection at astrophysical observations is a bit enhanced in the combined model with the `A' term of ${\cal O}(10)$ compared to those in the simplest left- and right-mediator models. The sensitivity of the present observations is, however, still not enough to put a constraint on the combined model. We therefore take into account the constraints from the direct mediator and WIMP production measurements, the Higgs and $Z$ boson branching fractions and the electroweak precision measurement at collider experiments based on those discussed in the previous sections in order to analyze the combined model.

\subsection{Motivation of the combined model}
\label{sec: g-2}

An important motivation for going beyond the simplest scenarios with left- or right- mediators and going to the combined one is from the anomalous muon magnetic moment\,\cite{Calibbi:2018rzv,Kawamura:2020qxo}, namely a long-standing discrepancy between the SM prediction and experimental results on the nature of the muon which may indicate the existence of new physics at the electroweak scale. It is also true that, as already mentioned at the beginning of section\,\ref{sec: model}, it is obligatory to introduce two types of the mediators regardless of the muon ($g - 2$) anomaly, as the left-handed and the right-handed fermions are different from the first beginning in the quantum field theory due to the chiral nature of the Lorentz symmetry in $(3+1)$ dimensional space-time. In other words, even if the muon ($g - 2$) anomaly disappears in the future, introducing the two types of the mediators is motivated to interpret experimental data correctly\,\cite{Ko:2016zxg}.

Going back to the muon ($g - 2$) anomaly, the discrepancy between the experimental result and the SM prediction reaches to 4.2$\sigma$ level, and is very recently reported to be $\Delta a_\mu \equiv a^\mathrm{exp}_\mu - a^\mathrm{SM}_\mu = 251(59) \times 10^{-11}$\,\cite{Abi:2021gix} at Fermilab. If the discrepancy persists even in future results by Fermilab\,\cite{Grange:2015fou} and also in J-PARC\,\cite{Iinuma:2011zz}, it is expected to become a smoking-gun signature of new physics beyond the SM at a certain low energy scale.

\begin{figure}[t]
    \centering
    \includegraphics[width=39mm]{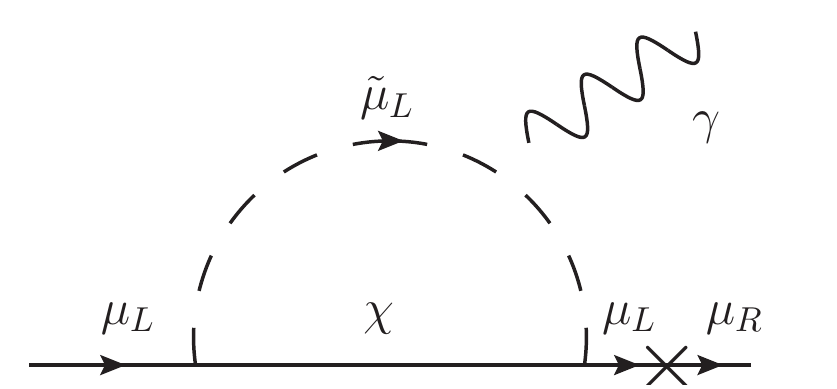}
    \includegraphics[width=39mm]{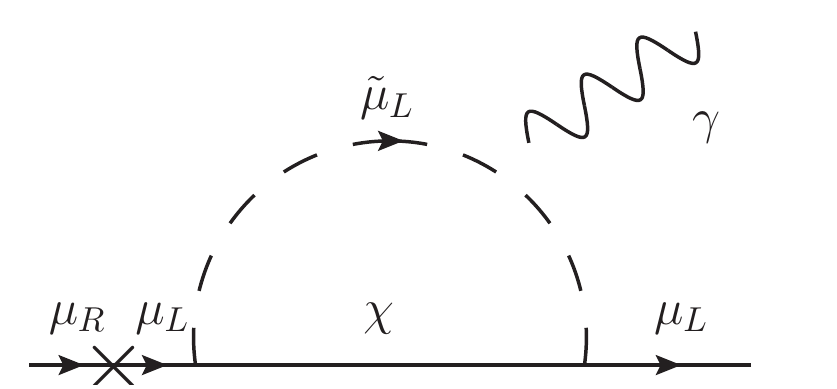}
    \includegraphics[width=39mm]{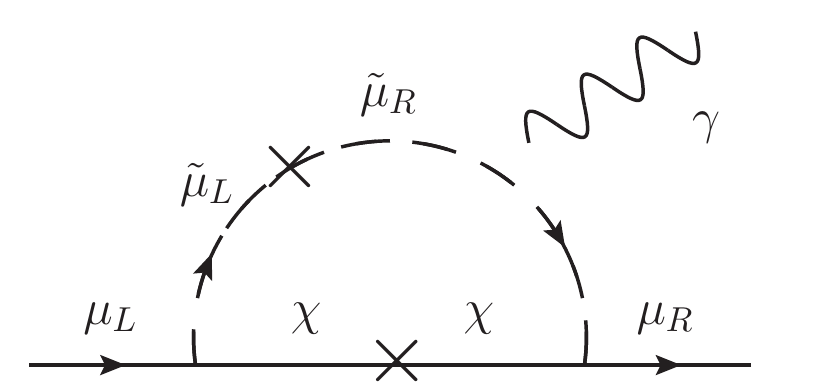} 
    \includegraphics[width=39mm]{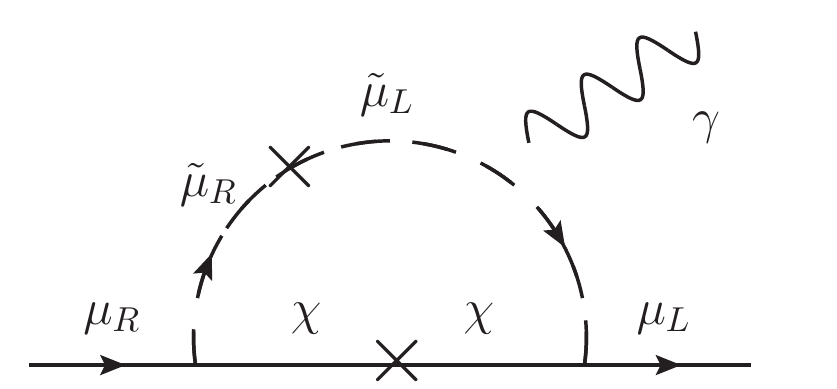}
    \caption{\small \sl Diagrams contributing to the anomalous muon magnetic moment. First two diagrams are those from the simplest leptophilic WIMP model with the left-handed mediators. Similar diagrams for the right-mediator model can be found by swapping $L$ with $R$. Last two diagrams are from the leptophilic WIMP model with both left- and right-handed mediators, namely the combined model.}
    \label{fig: g-2 diagrams}
\end{figure}

Since the leptophilic dark matter couples to the SM leptons, it potentially contributes to the anomalous magnetic moments to the leptons. In the case of the leptophilic dark matter only with left- or right-handed mediators, the contribution comes from the first two diagrams shown in Fig.\,\ref{fig: g-2 diagrams}. The explicit form of the contribution is given as follows\,\cite{Agrawal:2014ufa}:
\begin{align}
    & \Delta a^{\mathrm{L/R}}_\mu = -\frac{y_{L/R}^2}{96\pi^2} \frac{m_\mu^2}{m_{\tilde{\mu}_{L/R}}^2} f(r)
    \simeq - 10^{-10} \, y_{L/R}^2 \left(\frac{100\,\mathrm{GeV}}{m_{\tilde{\mu}_{L/R}}}\right)^2 f(r) ,
    \nonumber \\ 
    & f(r) = \frac{1 - 6r + 3r^2 + 2r^3 - 6r^2\log{r}}{(1-r)^4} ,
\end{align}
$\tilde{\mu}_{L/R} \equiv \tilde{e}_{L_2/R_2}$, $r \equiv m_\chi^2/m_{\tilde{\mu}_{L/R}}^2$, $f(r)$ is a monotonically decreasing function taking a value between zero and one. It is hard to explain the anomaly due to the negative contribution.

On the contrary, the combined model with left-handed and right-handed mediators yields an additional contribution to the anomalous magnetic moment via the last two diagrams in Fig.\,\ref{fig: g-2 diagrams}. Using the mass insertion approximation, its explicit form is given as\,\cite{Moroi:1995yh}
\begin{align}
    & \Delta a^{\mathrm{L+R}}_\mu = -\frac{y_L y_R}{96\pi^2} \frac{A\,v\,m_\mu^2}{\sqrt{2} m_\chi^3} f(x,y) ,
    \nonumber \\
    & f(x,y) = \frac{6}{xy}\left[\frac{-3+x+y+xy}{(x-1)^2(y-1)^2}+\frac{2x\log{x}}{(x-y)(x-1)^3}+\frac{2y\log{y}}{(y-x)(y-1)^3}\right] ,
    \label{eq: g-2 at conbined model}
\end{align}
where $x\equiv m^2_{\tilde{\mu}_{L}}/m^2_\chi$ and $y\equiv m^2_{\tilde{\mu}_{R}}/m^2_\chi$. The function $f(x,y)$ is a monotonically decreasing function (along with both $x$ and $y$ directions) taking a valued in between zero and one. Since the sign and size of the contribution $\Delta a^{\mathrm{L+R}}_\mu$ can be properly taken to account for the muon $(g-2)$ anomaly when we take an appropriate sign and value of the `A' term.

Since the function $f(x,y)$ becomes maximum taking the value of one when $x = y = 1$, the above contribution can be as large as $\Delta a^{\mathrm{L+R}}_\mu = -[y_L y_R/(96\pi^2)]\,[A v m_\mu^2/(\sqrt{2} m_\chi^3)] \simeq - 2 \times 10^{-9}\,y_L y_R\,A\,(100\,{\rm GeV}/m_\chi)^3$. Hence, the muon $(g-2)$ anomaly can always be explained even if the `A' term is ${\cal O}(1-10)$ whenever the dark matter as well as the mediators are at the electroweak scale. Conversely, it means that the electroweak scale WIMP and mediators, which are nothing but interesting targets at the first stage of the future lepton colliders, requires the size of the `A' term to be ${\cal O}(1-10)$ to explain the muon $(g-2)$ anomaly. 

Here, let us address the contribution of the leptophilic WIMP to the anomalous magnetic moment of the electron $\Delta a_e$. The contribution is given by the same formulae as above with replacing $m_\mu$ and $m_{\tilde{\mu}_{L/R}} = m_{\tilde{e}_{L_2/R_2}}$ with $m_e$ and $m_{\tilde{e}_{L_1/R_1}}$, respectively. As we have already mentioned in section\,\ref{subsec: g-2}, the contribution is negligibly small within the left- and right-mediator models. In the combined model, the contribution is predicted to be $\Delta a_e \simeq 6 \times 10^{-14}$ when we fix the value of the $A$ parameter so that it can explain the muon $(g-2)$ anomaly. Comparing this contribution to the experimental data shown in eq.\,(\ref{eq: electron g-2 discrepancy}), it turns out that its size is still within the error of the data. The observation of the anomalous magnetic moment of the electron hence does not put any severe constraints on the combined model.

\subsection{Present status and future prospects of the combined model}

\begin{figure}[t]
    \centering
    \includegraphics[width=73mm]{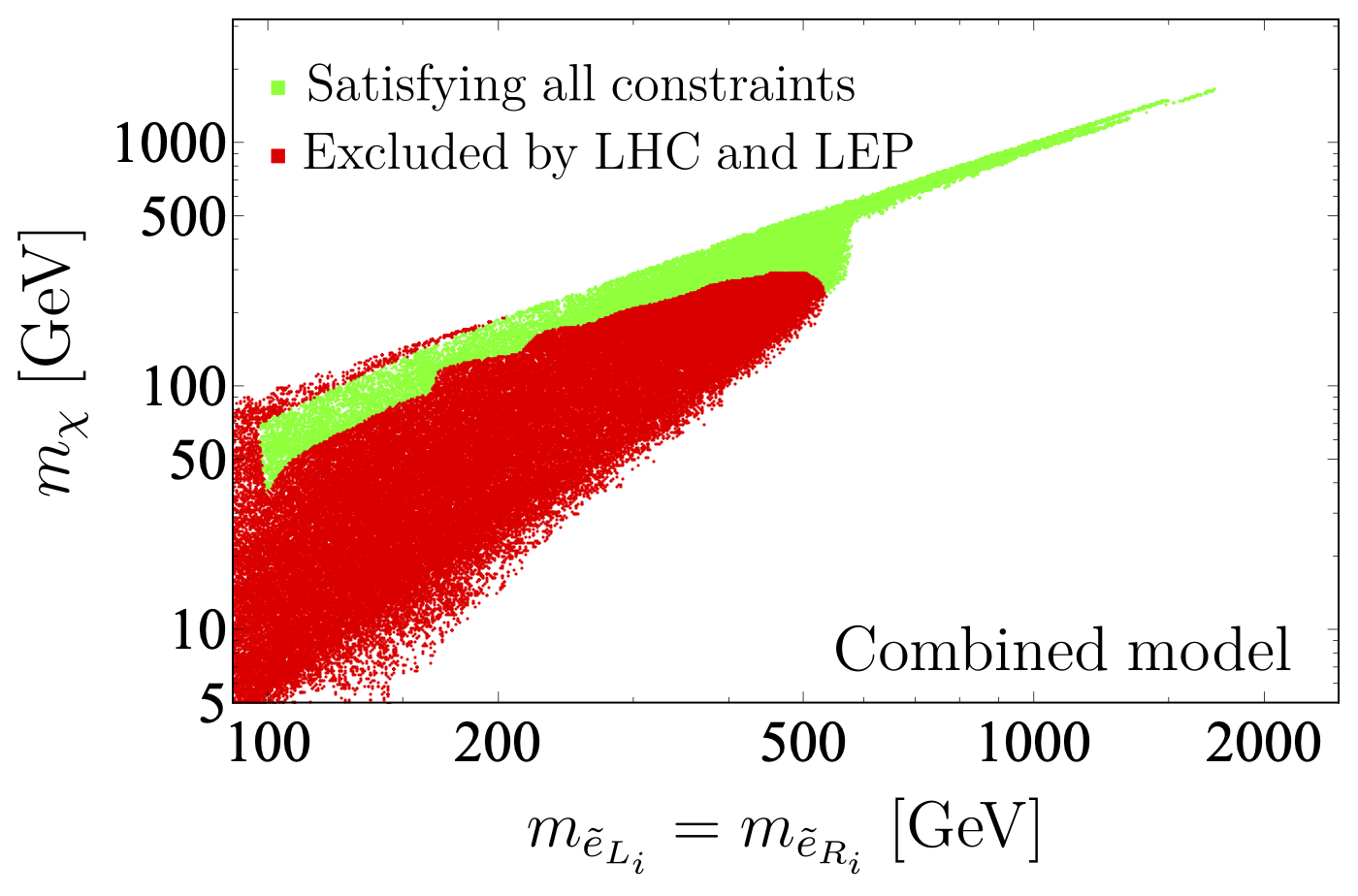}
    \qquad
    \includegraphics[width=73mm]{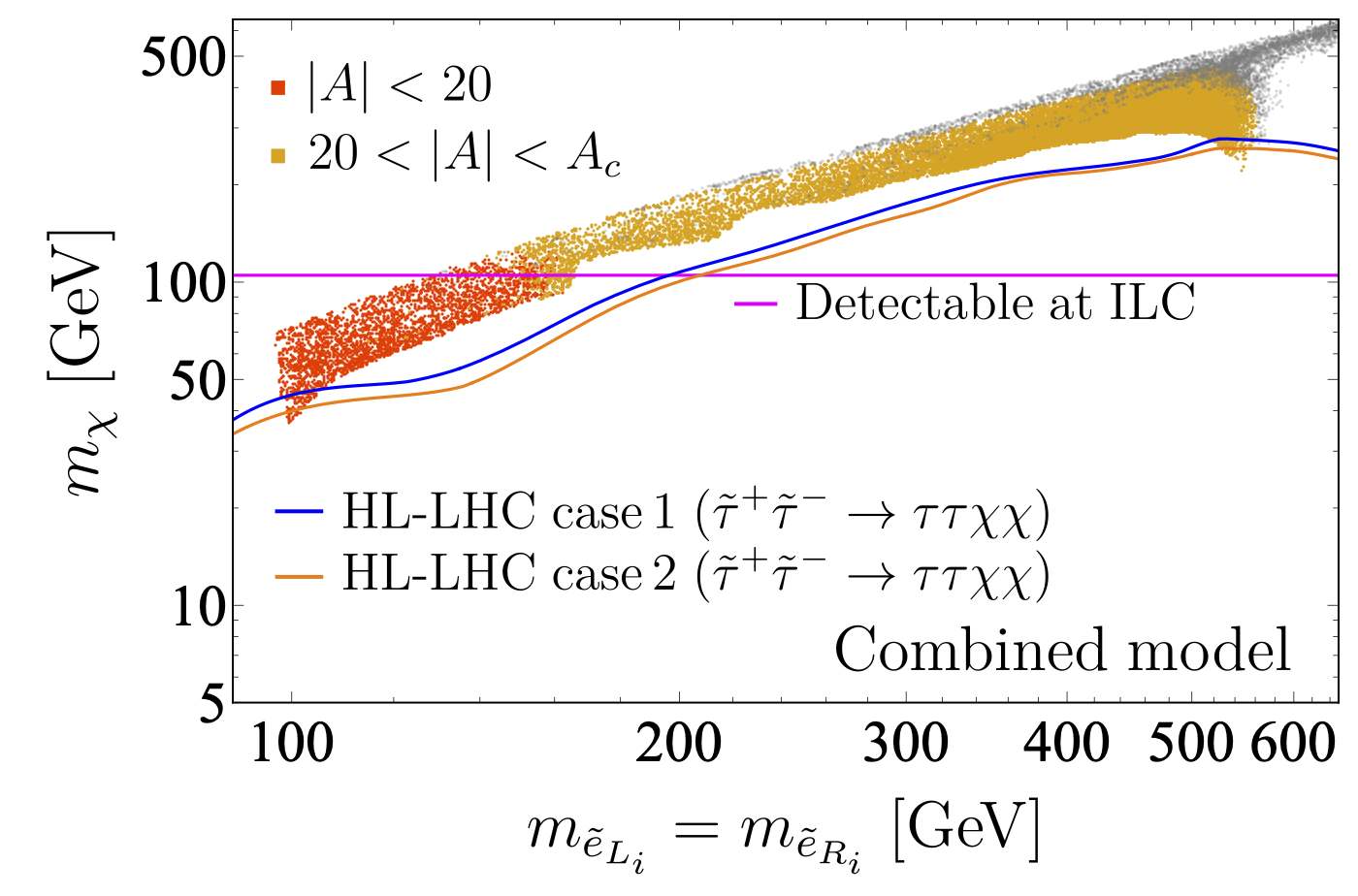} 
    \caption{\small \sl Allowed parameter space at 95\,\% C.L. for the combined model. The region spread by green points in the left panel is allowed by all the constraints discussed in section\,\ref{subsec: combined model}, while the one spread by red points is excluded only by the present LHC and LEP experiments. On the other hand, the region spread by all the points in the right panel is the same as the one spread by green points in the left panel with the color gradation indicating the value of the `A' term to explain the muon $(g-2)$ anomaly. $A_c$ is defined at section.\ref{subsubsec: VSC at combined model}. Projected 95\% C.L. reaches by the search for the direct mediator production at the future lepton collider (ILC) experiments are also shown by the magenta line. Blue and orange lines are expected 95\% C.L. reach of the direct stau pair production at the HL-LHC experiment based on two assumptions concerning systematic uncertainties.
}
    \label{fig: combined model}
\end{figure}

Present status and future prospects of the combined leptophilic WIMP model is shown in left and right panels of Fig.\,\ref{fig: combined model}, respectively, where the model parameters are scanned assuming $m_{\tilde{e}_L} = m_{\tilde{e}_R}$ and $y_R^4 = 2 y_L^4$, while imposing the constraints from the branching fraction measurements of Higgs and $Z$ boson decays and the electroweak precision measurements, as mentioned in  section\,\ref{subsec: combined model}. The color convention in the left panel is the same as those in Fig.\ref{fig: present status}. On the other hand, in the right panel, the region spread by all the points is the same as the one spread by green points in the left panel with the color gradation indicating the value of the `A' term to explain the muon $(g-2)$ anomaly ($\Delta a^{\mathrm{L/R}}_\mu = 251 \times 10^{-11}$). Here, $A_c$ is defined in section.\ref{subsubsec: VSC at combined model}. We note that some gray points for light mass region around $m_{\tilde{e}_{L_i}}=m_{\tilde{e}_{R_i}} \sim 300$\,GeV correspond to the co-annihilation of dark matter and sleptons, where $y_{L/R}$ can be very small. Moreover, the projected 95\% C.L. reach by the search for the direct WIMP or mediator productions at the future lepton collider (ILC) experiment is also shown by the magenta solid line, while expected 95\% C.L. reach of the direct stau pair production at the HL-LHC experiment based on two assumptions concerning systematic uncertainties are shown by the blue and orange lines. (See section.\ref{subsec: HL-LHC} for more detail.)

It is seen from the left panel of the figure that the allowed parameter region is similar to those of left- and right-handed mediator models discussed in the previous sections; The region with the large mass hierarchy between the WIMP and the mediators and that with the extremely degenerated mass hierarchy among the particles are excluded by the present LHC experiment, and the region that these particles are mildly degenerated is presently allowed at the electroweak mass scale of the WIMP. On the other hand, in the right panel of the figure, the future lepton collider (ILC) experiment is possible to explore the region with the `A' term less than ${\cal O}(10)$, and it verifies our discussion in section\,\ref{subsec: combined model}. We would like to further emphasize that such range of scalar mixing preserve the stability of our vacuum, i.e. $|A| < A_c$, as discussed in section\,\ref{subsubsec: VSC at combined model}. Hence, the future lepton colliders such as the ILC experiment is expected to play a complementary role to the HL-LHC experiment for the leptophilic WIMP and a unique role for the WIMP which explains the muon $(g-2)$ while being consistent with its thermal relic abundance to the dark matter density observed today.\footnote{In addition to the HL-LHC experiment, another experiment, the high-energy muon collider, is recently being discussed to search for the parameter region with heavier dark matter and mediator particles\,\cite{Capdevilla:2020qel, Capdevilla:2021rwo}.}

\section{Summary and Conclusions}
\label{sec: conclusion}

Minimal and renormalizable models for a singlet Majorana fermion WIMP that interacts only with the SM leptons via the scalar mediators are considered. We did an extensive analysis based on all robust theoretical and experimental constraints at present to show the feasibility of the models. To start with, we considered two distinct model scenarios; each has three generations of scalar mediators but the mediators in the first scenario are doublet under SM SU(2)$_L$ while the other scenario has only singlet mediators under the SU(2)$_L$ symmetry of the SM. Our choice of a Majorana fermion WIMP makes it necessary for the mediators to carry the exact same quantum numbers as their SM lepton partners. We considered the degenerate flavor-blind case for the sake of simplicity, and a single mass and a coupling parameter will define all the three generations of the scalar mediators.

The most important constraint comes from the relic abundance of the WIMP. We did a $\chi^2$-analysis to find out the allowed parameter space at 95\% C.L. to the experimental limit, but including additional theoretical uncertainties that come from SM thermodynamics at the early universe. The relic abundance constraint was further accompanied by theoretical one, such as vacuum stability, and electroweak precision data and all relevant direct search limits from the LEP and LHC run-II experiments. Our analysis show that at the large WIMP mass only coannihilation mechanism with the mediator particle can survive the relic abundance constraint and the largest allowed mass for the WIMP and the mediator can be around 1.5\,TeV. We also show that the direct search limit on the scalar mediators only discard a part of the parameter space that contributes to the relic abundance via the self-annihilation region for the WIMP mass below 300\,GeV and WIMP-mediator mass gap above 80\,GeV.

As a next step, we have discussed the possibility of probing the unexplored parameter region at future colliders, specifically at the future lepton colliders. We have done a signal-background analysis for the mono-photon search at the ILC experiment, and showed that the ILC-250\,GeV can indeed pin-down the WIMP mass around 110\,GeV with mildly degenerate WIMP-mediator mass. For the degenerate mass region, we have also included the pair production of the charged mediators\,\cite{Baum:2020gjj} as well as those of neutral mediators contributing non-negligibly to the mono-photon signal. Since it may be difficult to search for such a degenerate mass region at the HL-LHC experiment, the ILC experiment is expected to give a complementary search prospect to the future hadron collider for the leptophilic WIMP.

As the next to minimal extension, we have also considered the combined model scenario briefly where both doublet and singlet mediators contribute to the relic abundance. Introduction of a third degree of freedom to the WIMP co-annihilation decreases the largest allowed mass of the WIMP to 1.2 TeV. In the light of the anomalous magnetic moment of the muon, this combined model should be considered as the minimal scenario. Since, without the presence of both the mediator particles, it is impossible to explain the current discrepancy in the muon $(g-2)$. We show that we have ample parameter space that can simultaneously explain both the relic density and the muon $(g-2)$ anomaly. Moreover, we repeat the ILC mono-photon analysis for the combined model as well and found that the ILC-250 will be able to probe the aforementioned parameter space favored by the relic abundance and the muon $(g-2)$ data. If the muon anomaly persists further in the future experimental data then the ILC experiment will inevitably confirm or discard the model hypothesis.

\appendix

\section*{Acknowledgments}

This work is supported by Grant-in-Aid for Scientific Research from the Ministry of Education, Culture, Sports, Science, and Technology (MEXT), Japan; 17H02878, 20H01895, 19H05810, 20H00153 (for S. Matsumoto) and 18J21186 (for S. Horigome), by World Premier International Research Center Initiative (WPI), MEXT, Japan (Kavli IPMU), and also by JSPS Core-to-Core Program (JPJSCCA20200002) and in part by the National Science Foundation (Grant No. NSF PHY-1748958) through the Kavli Institute for Theoretical Physics (KITP) program “New Physics from Precision at High Energies”.

\bibliographystyle{JHEP}
\bibliography{refs}

\providecommand{\href}[2]{#2}\begingroup\raggedright\begin{thebibliography}{10}

\bibitem{Bernstein:1985th}
J.~Bernstein, L.~S. Brown and G.~Feinberg, \emph{{The Cosmological Heavy
  Neutrino Problem Revisited}},
  \href{http://dx.doi.org/10.1103/PhysRevD.32.3261}{\emph{Phys. Rev.} {\bf D32}
  (1985) 3261}.

\bibitem{Srednicki:1988ce}
M.~Srednicki, R.~Watkins and K.~A. Olive, \emph{{Calculations of Relic
  Densities in the Early Universe}},
  \href{http://dx.doi.org/10.1016/0550-3213(88)90099-5}{\emph{Nucl. Phys.} {\bf
  B310} (1988) 693}.

\bibitem{Boehm:2002yz}
C.~Boehm, T.~A. Ensslin and J.~Silk, \emph{{Can Annihilating dark matter be
  lighter than a few GeVs?}},
  \href{http://dx.doi.org/10.1088/0954-3899/30/3/004}{\emph{J. Phys.} {\bf G30}
  (2004) 279--286}, [\href{http://arxiv.org/abs/astro-ph/0208458}{{\tt
  astro-ph/0208458}}].

\bibitem{Boehm:2003bt}
C.~Boehm, D.~Hooper, J.~Silk, M.~Casse and J.~Paul, \emph{{MeV dark matter: Has
  it been detected?}},
  \href{http://dx.doi.org/10.1103/PhysRevLett.92.101301}{\emph{Phys. Rev.
  Lett.} {\bf 92} (2004) 101301},
  [\href{http://arxiv.org/abs/astro-ph/0309686}{{\tt astro-ph/0309686}}].

\bibitem{Griest:1989wd}
K.~Griest and M.~Kamionkowski, \emph{{Unitarity Limits on the Mass and Radius
  of Dark Matter Particles}},
  \href{http://dx.doi.org/10.1103/PhysRevLett.64.615}{\emph{Phys. Rev. Lett.}
  {\bf 64} (1990) 615}.

\bibitem{Hamaguchi:2007rb}
K.~Hamaguchi, S.~Shirai and T.~T. Yanagida, \emph{{Composite messenger baryon
  as a cold dark matter}},
  \href{http://dx.doi.org/10.1016/j.physletb.2007.08.047}{\emph{Phys. Lett.}
  {\bf B654} (2007) 110--112}, [\href{http://arxiv.org/abs/0707.2463}{{\tt
  0707.2463}}].

\bibitem{Hamaguchi:2008rv}
K.~Hamaguchi, E.~Nakamura, S.~Shirai and T.~T. Yanagida, \emph{{Decaying Dark
  Matter Baryons in a Composite Messenger Model}},
  \href{http://dx.doi.org/10.1016/j.physletb.2009.03.025}{\emph{Phys. Lett.}
  {\bf B674} (2009) 299--302}, [\href{http://arxiv.org/abs/0811.0737}{{\tt
  0811.0737}}].

\bibitem{Hamaguchi:2009db}
K.~Hamaguchi, E.~Nakamura, S.~Shirai and T.~T. Yanagida, \emph{{Low-Scale Gauge
  Mediation and Composite Messenger Dark Matter}},
  \href{http://dx.doi.org/10.1007/JHEP04(2010)119}{\emph{JHEP} {\bf 04} (2010)
  119}, [\href{http://arxiv.org/abs/0912.1683}{{\tt 0912.1683}}].

\bibitem{Murayama:2009nj}
H.~Murayama and J.~Shu, \emph{{Topological Dark Matter}},
  \href{http://dx.doi.org/10.1016/j.physletb.2010.02.037}{\emph{Phys. Lett.}
  {\bf B686} (2010) 162--165}, [\href{http://arxiv.org/abs/0905.1720}{{\tt
  0905.1720}}].

\bibitem{Hambye:2009fg}
T.~Hambye and M.~H.~G. Tytgat, \emph{{Confined hidden vector dark matter}},
  \href{http://dx.doi.org/10.1016/j.physletb.2009.11.050}{\emph{Phys. Lett.}
  {\bf B683} (2010) 39--41}, [\href{http://arxiv.org/abs/0907.1007}{{\tt
  0907.1007}}].

\bibitem{Antipin:2014qva}
O.~Antipin, M.~Redi and A.~Strumia, \emph{{Dynamical generation of the weak and
  Dark Matter scales from strong interactions}},
  \href{http://dx.doi.org/10.1007/JHEP01(2015)157}{\emph{JHEP} {\bf 01} (2015)
  157}, [\href{http://arxiv.org/abs/1410.1817}{{\tt 1410.1817}}].

\bibitem{Antipin:2015xia}
O.~Antipin, M.~Redi, A.~Strumia and E.~Vigiani, \emph{{Accidental Composite
  Dark Matter}}, \href{http://dx.doi.org/10.1007/JHEP07(2015)039}{\emph{JHEP}
  {\bf 07} (2015) 039}, [\href{http://arxiv.org/abs/1503.08749}{{\tt
  1503.08749}}].

\bibitem{Gross:2018zha}
C.~Gross, A.~Mitridate, M.~Redi, J.~Smirnov and A.~Strumia, \emph{{Cosmological
  Abundance of Colored Relics}},
  \href{http://dx.doi.org/10.1103/PhysRevD.99.016024}{\emph{Phys. Rev.} {\bf
  D99} (2019) 016024}, [\href{http://arxiv.org/abs/1811.08418}{{\tt
  1811.08418}}].

\bibitem{Fukuda:2018ufg}
H.~Fukuda, F.~Luo and S.~Shirai, \emph{{How Heavy can Neutralino Dark Matter
  be?}},  \href{http://arxiv.org/abs/1812.02066}{{\tt 1812.02066}}.

\bibitem{Dev:2013hka}
P.~S.~B. Dev, D.~K. Ghosh, N.~Okada and I.~Saha, \emph{{Neutrino Mass and Dark
  Matter in light of recent AMS-02 results}},
  \href{http://dx.doi.org/10.1103/PhysRevD.89.095001}{\emph{Phys. Rev. D} {\bf
  89} (2014) 095001}, [\href{http://arxiv.org/abs/1307.6204}{{\tt 1307.6204}}].

\bibitem{Bai:2014osa}
Y.~Bai and J.~Berger, \emph{{Lepton Portal Dark Matter}},
  \href{http://dx.doi.org/10.1007/JHEP08(2014)153}{\emph{JHEP} {\bf 08} (2014)
  153}, [\href{http://arxiv.org/abs/1402.6696}{{\tt 1402.6696}}].

\bibitem{Chang:2014tea}
S.~Chang, R.~Edezhath, J.~Hutchinson and M.~Luty, \emph{{Leptophilic Effective
  WIMPs}}, \href{http://dx.doi.org/10.1103/PhysRevD.90.015011}{\emph{Phys. Rev.
  D} {\bf 90} (2014) 015011}, [\href{http://arxiv.org/abs/1402.7358}{{\tt
  1402.7358}}].

\bibitem{Junius:2019dci}
S.~Junius, L.~Lopez-Honorez and A.~Mariotti, \emph{{A feeble window on
  leptophilic dark matter}},
  \href{http://dx.doi.org/10.1007/JHEP07(2019)136}{\emph{JHEP} {\bf 07} (2019)
  136}, [\href{http://arxiv.org/abs/1904.07513}{{\tt 1904.07513}}].

\bibitem{Okawa:2020jea}
S.~Okawa and Y.~Omura, \emph{{Light mass window of lepton portal dark matter}},
   \href{http://arxiv.org/abs/2011.04788}{{\tt 2011.04788}}.

\bibitem{Kawamura:2020qxo}
J.~Kawamura, S.~Okawa and Y.~Omura, \emph{{Current status and muon g-2
  explanation of lepton portal dark matter}},
  \href{http://dx.doi.org/10.1007/JHEP08(2020)042}{\emph{JHEP} {\bf 08} (2020)
  042}, [\href{http://arxiv.org/abs/2002.12534}{{\tt 2002.12534}}].

\bibitem{Banerjee:2016hsk}
S.~Banerjee, S.~Matsumoto, K.~Mukaida and Y.-L.~S. Tsai, \emph{{WIMP Dark
  Matter in a Well-Tempered Regime: A case study on Singlet-Doublets Fermionic
  WIMP}}, \href{http://dx.doi.org/10.1007/JHEP11(2016)070}{\emph{JHEP} {\bf 11}
  (2016) 070}, [\href{http://arxiv.org/abs/1603.07387}{{\tt 1603.07387}}].

\bibitem{Saikawa:2018rcs}
K.~Saikawa and S.~Shirai, \emph{{Primordial gravitational waves, precisely: The
  role of thermodynamics in the Standard Model}},
  \href{http://dx.doi.org/10.1088/1475-7516/2018/05/035}{\emph{JCAP} {\bf 1805}
  (2018) 035}, [\href{http://arxiv.org/abs/1803.01038}{{\tt 1803.01038}}].

\bibitem{Saikawa:2020swg}
K.~Saikawa and S.~Shirai, \emph{{Precise WIMP Dark Matter Abundance and
  Standard Model Thermodynamics}},
  \href{http://dx.doi.org/10.1088/1475-7516/2020/08/011}{\emph{JCAP} {\bf 08}
  (2020) 011}, [\href{http://arxiv.org/abs/2005.03544}{{\tt 2005.03544}}].

\bibitem{Jueid:2020yfj}
A.~Jueid, S.~Nasri and R.~Soualah, \emph{{Searching for GeV-scale Majorana Dark
  Matter: inter spem et metum}},  \href{http://arxiv.org/abs/2006.01348}{{\tt
  2006.01348}}.

\bibitem{Abi:2021gix}
{\scshape Muon g-2} collaboration, B.~Abi et~al., \emph{{Measurement of the
  Positive Muon Anomalous Magnetic Moment to 0.46~ppm}},
  \href{http://dx.doi.org/10.1103/PhysRevLett.126.141801}{\emph{Phys. Rev.
  Lett.} {\bf 126} (2021) 141801}, [\href{http://arxiv.org/abs/2104.03281}{{\tt
  2104.03281}}].

\bibitem{Ko:2016zxg}
P.~Ko, A.~Natale, M.~Park and H.~Yokoya, \emph{{Simplified DM models with the
  full SM gauge symmetry : the case of $t$-channel colored scalar mediators}},
  \href{http://dx.doi.org/10.1007/JHEP01(2017)086}{\emph{JHEP} {\bf 01} (2017)
  086}, [\href{http://arxiv.org/abs/1605.07058}{{\tt 1605.07058}}].

\bibitem{Madge:2018gfl}
E.~Madge and P.~Schwaller, \emph{{Leptophilic dark matter from gauged lepton
  number: Phenomenology and gravitational wave signatures}},
  \href{http://dx.doi.org/10.1007/JHEP02(2019)048}{\emph{JHEP} {\bf 02} (2019)
  048}, [\href{http://arxiv.org/abs/1809.09110}{{\tt 1809.09110}}].

\bibitem{Blanco:2019hah}
C.~Blanco, M.~Escudero, D.~Hooper and S.~J. Witte, \emph{{Z' mediated WIMPs:
  dead, dying, or soon to be detected?}},
  \href{http://dx.doi.org/10.1088/1475-7516/2019/11/024}{\emph{JCAP} {\bf 11}
  (2019) 024}, [\href{http://arxiv.org/abs/1907.05893}{{\tt 1907.05893}}].

\bibitem{Aghanim:2018eyx}
{\scshape Planck} collaboration, N.~Aghanim et~al., \emph{{Planck 2018 results.
  VI. Cosmological parameters}},  \href{http://arxiv.org/abs/1807.06209}{{\tt
  1807.06209}}.

\bibitem{Griest:1990kh}
K.~Griest and D.~Seckel, \emph{{Three exceptions in the calculation of relic
  abundances}}, \href{http://dx.doi.org/10.1103/PhysRevD.43.3191}{\emph{Phys.
  Rev. D} {\bf 43} (1991) 3191--3203}.

\bibitem{Belanger:2018mqt}
G.~Bélanger, F.~Boudjema, A.~Goudelis, A.~Pukhov and B.~Zaldivar,
  \emph{{micrOMEGAs5.0 : Freeze-in}},
  \href{http://dx.doi.org/10.1016/j.cpc.2018.04.027}{\emph{Comput. Phys.
  Commun.} {\bf 231} (2018) 173--186},
  [\href{http://arxiv.org/abs/1801.03509}{{\tt 1801.03509}}].

\bibitem{Baker:2018uox}
M.~J. Baker and A.~Thamm, \emph{{Leptonic WIMP Coannihilation and the Current
  Dark Matter Search Strategy}},
  \href{http://dx.doi.org/10.1007/JHEP10(2018)187}{\emph{JHEP} {\bf 10} (2018)
  187}, [\href{http://arxiv.org/abs/1806.07896}{{\tt 1806.07896}}].

\bibitem{LEPII}
{\scshape ALEPH, DELPHI, L3, OPAL Experiments} collaboration
  \href{http://arxiv.org/abs/http://lepsusy.web.cern.ch/lepsusy/}{{\tt
  http://lepsusy.web.cern.ch/lepsusy/}}.

\bibitem{pdg2020}
P.~D.~G. 2020, \emph{{Review of Particle Physics}},
  \href{http://dx.doi.org/10.1093/ptep/ptaa104}{\emph{{Progress of Theoretical
  and Experimental Physics}} {\bf 2020} (08, 2020) }.

\bibitem{Zyla:2020zbs}
{\scshape Particle Data Group} collaboration, P.~Zyla et~al., \emph{{Review of
  Particle Physics}}, \href{http://dx.doi.org/10.1093/ptep/ptaa104}{\emph{PTEP}
  {\bf 2020} (2020) 083C01}.

\bibitem{Datta:2002jh}
A.~Datta and A.~Datta, \emph{{Are light sneutrinos buried in LEP data?}},
  \href{http://dx.doi.org/10.1016/j.physletb.2003.10.023}{\emph{Phys. Lett. B}
  {\bf 578} (2004) 165--175}, [\href{http://arxiv.org/abs/hep-ph/0210218}{{\tt
  hep-ph/0210218}}].

\bibitem{Aad:2019vnb}
{\scshape ATLAS} collaboration, G.~Aad et~al., \emph{{Search for electroweak
  production of charginos and sleptons decaying into final states with two
  leptons and missing transverse momentum in $\sqrt{s}=13$ TeV $pp$ collisions
  using the ATLAS detector}},
  \href{http://dx.doi.org/10.1140/epjc/s10052-019-7594-6}{\emph{Eur. Phys. J.
  C} {\bf 80} (2020) 123}, [\href{http://arxiv.org/abs/1908.08215}{{\tt
  1908.08215}}].

\bibitem{Aad:2019qnd}
{\scshape ATLAS} collaboration, G.~Aad et~al., \emph{{Searches for electroweak
  production of supersymmetric particles with compressed mass spectra in
  $\sqrt{s}=$ 13 TeV $pp$ collisions with the ATLAS detector}},
  \href{http://dx.doi.org/10.1103/PhysRevD.101.052005}{\emph{Phys. Rev. D} {\bf
  101} (2020) 052005}, [\href{http://arxiv.org/abs/1911.12606}{{\tt
  1911.12606}}].

\bibitem{Aad:2019byo}
{\scshape ATLAS} collaboration, G.~Aad et~al., \emph{{Search for direct stau
  production in events with two hadronic $\tau$-leptons in $\sqrt{s} = 13$ TeV
  $pp$ collisions with the ATLAS detector}},
  \href{http://dx.doi.org/10.1103/PhysRevD.101.032009}{\emph{Phys. Rev. D} {\bf
  101} (2020) 032009}, [\href{http://arxiv.org/abs/1911.06660}{{\tt
  1911.06660}}].

\bibitem{Sirunyan:2019mlu}
{\scshape CMS} collaboration, A.~M. Sirunyan et~al., \emph{{Search for
  Supersymmetry with a Compressed Mass Spectrum in Events with a Soft $\tau$
  Lepton, a Highly Energetic Jet, and Large Missing Transverse Momentum in
  Proton-Proton Collisions at $\sqrt{s}=$ TeV}},
  \href{http://dx.doi.org/10.1103/PhysRevLett.124.041803}{\emph{Phys. Rev.
  Lett.} {\bf 124} (2020) 041803}, [\href{http://arxiv.org/abs/1910.01185}{{\tt
  1910.01185}}].

\bibitem{Balazs:2017ple}
J.~Ellis, A.~Fowlie, L.~Marzola and M.~Raidal, \emph{{Statistical Analyses of
  Higgs- and Z-Portal Dark Matter Models}},
  \href{http://dx.doi.org/10.1103/PhysRevD.97.115014}{\emph{Phys. Rev. D} {\bf
  97} (2018) 115014}, [\href{http://arxiv.org/abs/1711.09912}{{\tt
  1711.09912}}].

\bibitem{Gunion:1989we}
J.~F. Gunion, H.~E. Haber, G.~L. Kane and S.~Dawson, \emph{{The Higgs Hunter's
  Guide}}, vol.~80.
\newblock 2000.

\bibitem{Djouadi:2005gi}
A.~Djouadi, \emph{{The Anatomy of electro-weak symmetry breaking. I: The Higgs
  boson in the standard model}},
  \href{http://dx.doi.org/10.1016/j.physrep.2007.10.004}{\emph{Phys. Rept.}
  {\bf 457} (2008) 1--216}, [\href{http://arxiv.org/abs/hep-ph/0503172}{{\tt
  hep-ph/0503172}}].

\bibitem{Djouadi:2005gj}
A.~Djouadi, \emph{{The Anatomy of electro-weak symmetry breaking. II. The Higgs
  bosons in the minimal supersymmetric model}},
  \href{http://dx.doi.org/10.1016/j.physrep.2007.10.005}{\emph{Phys. Rept.}
  {\bf 459} (2008) 1--241}, [\href{http://arxiv.org/abs/hep-ph/0503173}{{\tt
  hep-ph/0503173}}].

\bibitem{Sirunyan:2018ouh}
{\scshape CMS} collaboration, A.~Sirunyan et~al., \emph{{Measurements of Higgs
  boson properties in the diphoton decay channel in proton-proton collisions at
  $\sqrt{s} =$ 13 TeV}},
  \href{http://dx.doi.org/10.1007/JHEP11(2018)185}{\emph{JHEP} {\bf 11} (2018)
  185}, [\href{http://arxiv.org/abs/1804.02716}{{\tt 1804.02716}}].

\bibitem{Djouadi:1996pb}
A.~Djouadi, V.~Driesen, W.~Hollik and J.~I. Illana, \emph{{The Coupling of the
  lightest SUSY Higgs boson to two photons in the decoupling regime}},
  \href{http://dx.doi.org/10.1007/BF01245805}{\emph{Eur. Phys. J. C} {\bf 1}
  (1998) 149--162}, [\href{http://arxiv.org/abs/hep-ph/9612362}{{\tt
  hep-ph/9612362}}].

\bibitem{He:2001tp}
H.-J. He, N.~Polonsky and S.-f. Su, \emph{{Extra families, Higgs spectrum and
  oblique corrections}},
  \href{http://dx.doi.org/10.1103/PhysRevD.64.053004}{\emph{Phys. Rev. D} {\bf
  64} (2001) 053004}, [\href{http://arxiv.org/abs/hep-ph/0102144}{{\tt
  hep-ph/0102144}}].

\bibitem{Grimus:2007if}
W.~Grimus, L.~Lavoura, O.~Ogreid and P.~Osland, \emph{{A Precision constraint
  on multi-Higgs-doublet models}},
  \href{http://dx.doi.org/10.1088/0954-3899/35/7/075001}{\emph{J. Phys. G} {\bf
  35} (2008) 075001}, [\href{http://arxiv.org/abs/0711.4022}{{\tt 0711.4022}}].

\bibitem{Grimus:2008nb}
W.~Grimus, L.~Lavoura, O.~Ogreid and P.~Osland, \emph{{The Oblique parameters
  in multi-Higgs-doublet models}},
  \href{http://dx.doi.org/10.1016/j.nuclphysb.2008.04.019}{\emph{Nucl. Phys. B}
  {\bf 801} (2008) 81--96}, [\href{http://arxiv.org/abs/0802.4353}{{\tt
  0802.4353}}].

\bibitem{Barbieri:2006dq}
R.~Barbieri, L.~J. Hall and V.~S. Rychkov, \emph{{Improved naturalness with a
  heavy Higgs: An Alternative road to LHC physics}},
  \href{http://dx.doi.org/10.1103/PhysRevD.74.015007}{\emph{Phys. Rev. D} {\bf
  74} (2006) 015007}, [\href{http://arxiv.org/abs/hep-ph/0603188}{{\tt
  hep-ph/0603188}}].

\bibitem{ALEPH:2005ab}
{\scshape ALEPH, DELPHI, L3, OPAL, SLD, LEP Electroweak Working Group, SLD
  Electroweak Group, SLD Heavy Flavour Group} collaboration, S.~Schael et~al.,
  \emph{{Precision electroweak measurements on the $Z$ resonance}},
  \href{http://dx.doi.org/10.1016/j.physrep.2005.12.006}{\emph{Phys. Rept.}
  {\bf 427} (2006) 257--454}, [\href{http://arxiv.org/abs/hep-ex/0509008}{{\tt
  hep-ex/0509008}}].

\bibitem{Asner:2013psa}
D.~M. Asner et~al., \emph{{ILC Higgs White Paper}},  in \emph{{Community Summer
  Study 2013}: {Snowmass on the Mississippi}}, 10, 2013.
\newblock \href{http://arxiv.org/abs/1310.0763}{{\tt 1310.0763}}.

\bibitem{ATLAS:2020cjb}
{\scshape ATLAS} collaboration, \emph{{Search for invisible Higgs boson decays
  with vector boson fusion signatures with the ATLAS detector using an
  integrated luminosity of 139 fb$^{-1}$}}, .

\bibitem{Hanneke:2008tm}
D.~Hanneke, S.~Fogwell and G.~Gabrielse, \emph{{New Measurement of the Electron
  Magnetic Moment and the Fine Structure Constant}},
  \href{http://dx.doi.org/10.1103/PhysRevLett.100.120801}{\emph{Phys. Rev.
  Lett.} {\bf 100} (2008) 120801}, [\href{http://arxiv.org/abs/0801.1134}{{\tt
  0801.1134}}].

\bibitem{Parker_2018}
R.~H. Parker, C.~Yu, W.~Zhong, B.~Estey and H.~Müller, \emph{Measurement of
  the fine-structure constant as a test of the standard model},
  \href{http://dx.doi.org/10.1126/science.aap7706}{\emph{Science} {\bf 360}
  (Apr, 2018) 191–195}.

\bibitem{2013PASP..125..306F}
D.~{Foreman-Mackey}, D.~W. {Hogg}, D.~{Lang} and J.~{Goodman}, \emph{{emcee:
  The MCMC Hammer}}, \href{http://dx.doi.org/10.1086/670067}{\emph{Publications
  of the Astronomical Society of the Pacific} {\bf 125} (Mar., 2013) 306},
  [\href{http://arxiv.org/abs/1202.3665}{{\tt 1202.3665}}].

\bibitem{Cepeda:2019klc}
M.~Cepeda et~al., \emph{{Report from Working Group 2}: {Higgs Physics at the
  HL-LHC and HE-LHC}}, vol.~7, pp.~221--584.
\newblock 12, 2019.
\newblock \href{http://arxiv.org/abs/1902.00134}{{\tt 1902.00134}}.
\newblock 10.23731/CYRM-2019-007.221.

\bibitem{deBlas:2019rxi}
J.~de~Blas et~al., \emph{{Higgs Boson Studies at Future Particle Colliders}},
  \href{http://dx.doi.org/10.1007/JHEP01(2020)139}{\emph{JHEP} {\bf 01} (2020)
  139}, [\href{http://arxiv.org/abs/1905.03764}{{\tt 1905.03764}}].

\bibitem{ATLAS:2018diz}
{\scshape ATLAS} collaboration, \emph{{Prospects for searches for staus,
  charginos and neutralinos at the high luminosity LHC with the ATLAS
  Detector}}, .

\bibitem{Baum:2020gjj}
S.~Baum, P.~Sandick and P.~Stengel, \emph{{Hunting for scalar lepton partners
  at future electron colliders}},
  \href{http://dx.doi.org/10.1103/PhysRevD.102.015026}{\emph{Phys. Rev. D} {\bf
  102} (2020) 015026}, [\href{http://arxiv.org/abs/2004.02834}{{\tt
  2004.02834}}].

\bibitem{Ghosh:2019rtj}
D.~K. Ghosh, T.~Katayose, S.~Matsumoto, I.~Saha, S.~Shirai and T.~Tanabe,
  \emph{{Role of future lepton colliders for fermionic $Z$-portal dark matter
  models}}, \href{http://dx.doi.org/10.1103/PhysRevD.101.015007}{\emph{Phys.
  Rev. D} {\bf 101} (2020) 015007},
  [\href{http://arxiv.org/abs/1906.06864}{{\tt 1906.06864}}].

\bibitem{Fox:2011fx}
P.~J. Fox, R.~Harnik, J.~Kopp and Y.~Tsai, \emph{{LEP Shines Light on Dark
  Matter}}, \href{http://dx.doi.org/10.1103/PhysRevD.84.014028}{\emph{Phys.
  Rev.} {\bf D84} (2011) 014028}, [\href{http://arxiv.org/abs/1103.0240}{{\tt
  1103.0240}}].

\bibitem{Kuraev:1985hb}
E.~A. Kuraev and V.~S. Fadin, \emph{{On Radiative Corrections to e+ e- Single
  Photon Annihilation at High-Energy}}, {\emph{Sov. J. Nucl. Phys.} {\bf 41}
  (1985) 466--472}.

\bibitem{Datta:2005gm}
A.~K. Datta, K.~Kong and K.~T. Matchev, \emph{{The Impact of beamstrahlung on
  precision measurements at CLIC}}, {\emph{eConf} {\bf C050318} (2005) 0215},
  [\href{http://arxiv.org/abs/hep-ph/0508161}{{\tt hep-ph/0508161}}].

\bibitem{Behnke:2013xla}
T.~Behnke, J.~E. Brau, B.~Foster, J.~Fuster, M.~Harrison, J.~M. Paterson
  et~al., \emph{{The International Linear Collider Technical Design Report -
  Volume 1: Executive Summary}},  \href{http://arxiv.org/abs/1306.6327}{{\tt
  1306.6327}}.

\bibitem{DAmbrosio:2002vsn}
G.~D'Ambrosio, G.~F. Giudice, G.~Isidori and A.~Strumia, \emph{{Minimal flavor
  violation: An Effective field theory approach}},
  \href{http://dx.doi.org/10.1016/S0550-3213(02)00836-2}{\emph{Nucl. Phys. B}
  {\bf 645} (2002) 155--187}, [\href{http://arxiv.org/abs/hep-ph/0207036}{{\tt
  hep-ph/0207036}}].

\bibitem{Colangelo:2008qp}
G.~Colangelo, E.~Nikolidakis and C.~Smith, \emph{{Supersymmetric models with
  minimal flavour violation and their running}},
  \href{http://dx.doi.org/10.1140/epjc/s10052-008-0796-y}{\emph{Eur. Phys. J.
  C} {\bf 59} (2009) 75--98}, [\href{http://arxiv.org/abs/0807.0801}{{\tt
  0807.0801}}].

\bibitem{Duan:2018cgb}
G.~H. Duan, C.~Han, B.~Peng, L.~Wu and J.~M. Yang, \emph{{Vacuum stability in
  stau-neutralino coannihilation in MSSM}},
  \href{http://dx.doi.org/10.1016/j.physletb.2018.12.001}{\emph{Phys. Lett. B}
  {\bf 788} (2019) 475--479}, [\href{http://arxiv.org/abs/1809.10061}{{\tt
  1809.10061}}].

\bibitem{Hiller:2019mou}
G.~Hiller, C.~Hormigos-Feliu, D.~F. Litim and T.~Steudtner, \emph{{Anomalous
  magnetic moments from asymptotic safety}},
  \href{http://dx.doi.org/10.1103/PhysRevD.102.071901}{\emph{Phys. Rev. D} {\bf
  102} (2020) 071901}, [\href{http://arxiv.org/abs/1910.14062}{{\tt
  1910.14062}}].

\bibitem{Hiller:2020fbu}
G.~Hiller, C.~Hormigos-Feliu, D.~F. Litim and T.~Steudtner, \emph{{Model
  Building from Asymptotic Safety with Higgs and Flavor Portals}},
  \href{http://dx.doi.org/10.1103/PhysRevD.102.095023}{\emph{Phys. Rev. D} {\bf
  102} (2020) 095023}, [\href{http://arxiv.org/abs/2008.08606}{{\tt
  2008.08606}}].

\bibitem{Calibbi:2018rzv}
L.~Calibbi, R.~Ziegler and J.~Zupan, \emph{{Minimal models for dark matter and
  the muon g\ensuremath{-}2 anomaly}},
  \href{http://dx.doi.org/10.1007/JHEP07(2018)046}{\emph{JHEP} {\bf 07} (2018)
  046}, [\href{http://arxiv.org/abs/1804.00009}{{\tt 1804.00009}}].

\bibitem{Grange:2015fou}
{\scshape Muon g-2} collaboration, J.~Grange et~al., \emph{{Muon (g-2)
  Technical Design Report}},  \href{http://arxiv.org/abs/1501.06858}{{\tt
  1501.06858}}.

\bibitem{Iinuma:2011zz}
{\scshape J-PARC muon g-2/EDM} collaboration, H.~Iinuma, \emph{{New approach to
  the muon g-2 and EDM experiment at J-PARC}},
  \href{http://dx.doi.org/10.1088/1742-6596/295/1/012032}{\emph{J. Phys. Conf.
  Ser.} {\bf 295} (2011) 012032}.

\bibitem{Agrawal:2014ufa}
P.~Agrawal, Z.~Chacko and C.~B. Verhaaren, \emph{{Leptophilic Dark Matter and
  the Anomalous Magnetic Moment of the Muon}},
  \href{http://dx.doi.org/10.1007/JHEP08(2014)147}{\emph{JHEP} {\bf 08} (2014)
  147}, [\href{http://arxiv.org/abs/1402.7369}{{\tt 1402.7369}}].

\bibitem{Moroi:1995yh}
T.~Moroi, \emph{{The Muon anomalous magnetic dipole moment in the minimal
  supersymmetric standard model}},
  \href{http://dx.doi.org/10.1103/PhysRevD.53.6565}{\emph{Phys. Rev. D} {\bf
  53} (1996) 6565--6575}, [\href{http://arxiv.org/abs/hep-ph/9512396}{{\tt
  hep-ph/9512396}}].

\bibitem{Capdevilla:2020qel}
R.~Capdevilla, D.~Curtin, Y.~Kahn and G.~Krnjaic, \emph{{A Guaranteed Discovery
  at Future Muon Colliders}},  \href{http://arxiv.org/abs/2006.16277}{{\tt
  2006.16277}}.

\bibitem{Capdevilla:2021rwo}
R.~Capdevilla, D.~Curtin, Y.~Kahn and G.~Krnjaic, \emph{{A No-Lose Theorem for
  Discovering the New Physics of $(g-2)_\mu$ at Muon Colliders}},
  \href{http://arxiv.org/abs/2101.10334}{{\tt 2101.10334}}.

\end{thebibliography}\endgroup

\end{document}